\documentclass[11pt,a4paper]{scrartcl} 

\usepackage[utf8]{inputenc}
\usepackage[T1]{fontenc}
\usepackage{lmodern}

\usepackage{graphicx}
\usepackage{epsfig}
\usepackage{float}
\usepackage{cite}
\usepackage{amssymb}
\usepackage{amsmath}
\usepackage{url}
\usepackage[dvipsnames]{xcolor}
\usepackage{cancel}
\usepackage{nicefrac}
\usepackage[normalem]{ulem}
\usepackage[toc,page]{appendix}
\usepackage[onehalfspacing]{setspace}
\usepackage{booktabs}

\usepackage[format=plain,labelfont={bf},font={small}]{caption}

\usepackage{hyperref}

\usepackage[tight]{subfigure}

\usepackage{comment}

\usepackage{tikz,pgfplots}
\usetikzlibrary{decorations.pathmorphing, patterns}
\usepackage[compat=1.1.0]{tikz-feynman}

\setkomafont{disposition}{\normalcolor\normalfont\bfseries}

\newcommand*{\alphas}{\alpha_{\text{s}}}

\newcommand*{\refeq}[1]{Eq.~\eqref{#1}}

\newcommand*{\mr}{\mathrm}

\newcommand*{\beq}{\begin{equation}}
\newcommand*{\eeq}{\end{equation}}
\newcommand*{\bea}{\begin{eqnarray}}
\newcommand*{\eea}{\end{eqnarray}}

\newcommand{\gev}{\mathrm{GeV}}

\newcommand*{\bi}{\begin{itemize}}
\newcommand*{\ei}{\end{itemize}}

\newcommand*{\muf}{\mu_\mr{F}} 
\newcommand*{\mur}{\mu_\mr{R}} 
\newcommand*{\xif}{\xi_\mr{F}} 
\newcommand*{\xir}{\xi_\mr{R}}

\newcommand{\order}[1]{\mathcal{O}(#1)}

\newcommand{\ystar}{y_{j}^*} 
\newcommand{\ystart}{y_{j_3}^*} 
\newcommand{\ystarf}{y_{j_4}^*} 

\newcommand*{\PBOX}{\texttt{POWHEG-BOX}} 

\newcommand*{\PBOXRES}{\texttt{POWHEG-BOX-RES}}

\newcommand*{\POWHEG}{\texttt{POWHEG}}

\newcommand*{\PYTHIA}{\texttt{PYTHIA}} 
\newcommand*{\PYTHIAE}{\texttt{PYTHIA~8}}

\newcommand*{\PYDS}{\texttt{PY+DS}} 
\newcommand*{\PYGS}{\texttt{PY+GS}} 
\newcommand*{\PYMPI}{\texttt{PY+DS+MPI+HAD}}

\newcommand*{\MG}{\texttt{MadGraph5\_aMC@NLO}} 
\newcommand*{\Madgraph}{\texttt{MadGraph}}

\newcommand*{\recola}{\texttt{RECOLA}}
\newcommand*{\recolatwo}{\texttt{RECOLA2}}

\newcommand{\wpwpjjj}{W^+W^+jjj}
\newcommand{\wpwp}{W^+W^+}

\newcommand{\wpwpjj}{W^+W^+jj}

\newcommand{\ptveto}{p_T^\mr{veto}}

\newcommand{\NLOPS}{NLO+PS}

\definecolor{dodgerblue}{rgb}{0.12, 0.56, 1.0}

\definecolor{darkgreen}{rgb}{0.0, 0.5, 0.0}
\definecolor{teal}{rgb}{0.0, 0.5, 0.5}          % Teal
\definecolor{purple}{rgb}{0.5, 0.0, 0.5}        % Purple

\definecolor{mypink}{rgb}{0.91, 0.33, 0.51}

\newcommand{\ed}{\end{document}}

\begin{document}

\renewcommand*{\thefootnote}{\fnsymbol{footnote}}

\begin{center}
	{
\Large \textbf{Electroweak $\wpwp$ production in association with \\[1ex]
three jets at NLO QCD matched with parton shower}}\\
	\vspace{.7cm}
	Barbara J\"ager\footnote{\texttt{jaeger@itp.uni-tuebingen.de}}, 
	\\
	\textit{
		Institute for Theoretical Physics, University of T\"ubingen, Auf der Morgenstelle 14, 72076 T\"ubingen, Germany
	}
	\\[.3cm]		 
	Santiago Lopez Portillo Chavez\footnote{\texttt{santiago.lopez-portillo-chavez@uni-wuerzburg.de}},
	\\
	\textit{
		Institute for Theoretical Physics and Astrophysics, University of Würzburg, Emil-Hilb-Weg 22, 97074 Würzburg, Germany
	}
\end{center}   

\renewcommand*{\thefootnote}{\arabic{footnote}}
\setcounter{footnote}{0}

\vspace*{0.1cm}
\begin{abstract} 
We present a next-to-leading order QCD calculation for the electroweak production of a pair of same-sign weak gauge bosons in association with three jets in the region dominated by vector-boson scattering, and its  matching with parton-shower programs according to the \POWHEG{} formalism. Our calculation allows for an accurate description of jet observables crucial for the optimization of experimental analyses of vector boson scattering processes. 
We find QCD corrections and parton-shower effects moderate in size for inclusive quantities and distributions of the tagging jets. A larger impact of these corrections is observed for subleading jets. 
\end{abstract}
\newpage

\tableofcontents

%%%%%%%%%%%%%%%%%%%%%%%%%%%%%%%%%%%%%%%%%%%
\section{Introduction}\label{sec:intro}
%%%%%%%%%%%%%%%%%%%%%%%%%%%%%%%%%%%%%%%%%%%
%
Vector-boson scattering (VBS) processes represent an important window into the electroweak (EW) sector of the Standard Model (SM), 
being sensitive to triple and quartic EW gauge couplings, as well as to couplings of the Higgs boson to EW vector bosons. 
Correspondingly, large efforts have been made to investigate these processes both experimentally and theoretically. 

In 2014, the first hint of same-sign $W$-boson scattering at the LHC in the fully leptonic channel was reported by the ATLAS collaboration~\cite{ATLAS:2014jzl}. 
In the following years, its observation was confirmed by both the CMS~\cite{CMS:2014mra,CMS:2017fhs} 
and ATLAS~\cite{ATLAS:2016snd,ATLAS:2019cbr,ATLAS:2023sua} collaborations, as well as the observation of processes involving the scattering of
$ZZ$~\cite{CMS:2017zmo,ATLAS:2020nlt,CMS:2020fqz,ATLAS:2023dkz}, $WZ$~\cite{ATLAS:2018mxa,CMS:2019uys,CMS:2020gfh,ATLAS:2024ini}, 
and $W^+W^-$ bosons~\cite{CMS:2022woe,ATLAS:2024ett}.

Calculations of massive vector-boson scattering processes at leading order (LO) have been available for around 30 years~\cite{Barger:1990py,Barger:1991ar}. 
Corrections at next-to-leading order (NLO) in QCD have been calculated for the scattering of $W^+W^-$, $ZZ$, $W^\pm W^\pm$ and $W^\pm Z$ in Refs.~\cite{Jager:2006zc,Jager:2006cp,Jager:2009xx,Bozzi:2007ur,Denner:2012dz} and have subsequently been matched to a parton shower (PS)~\cite{Jager:2013mu,Jager:2013iza,Jager:2011ms,Jager:2018cyo,Jager:2024sij}. These calculations rely on the so-called VBS approximation, where $s$-channel diagrams as well as the interference of $t$- and $u$-channel diagrams are neglected.

More recently, exact NLO EW corrections to $W^\pm W^\pm$ scattering were presented~\cite{Biedermann:2016yds}, and subsequently the complete NLO corrections, which include EW, QCD and mixed contributions~\cite{Biedermann:2017bss,Dittmaier:2023nac}. 
The quality of the VBS approximation was analysed in Refs.~\cite{Ballestrero:2018anz,Dittmaier:2023nac} for the $W^+W^+$ mode 
and in Ref.~\cite{Campanario:2018ppz} for the related process of $Hjj$ production via vector-boson fusion (VBF). There, it was found that the approximation is justified in certain regions of the phase space. In Ref.~\cite{Chiesa:2019ulk}, the NLO EW corrections to $W^\pm W^\pm$ scattering were matched with parton showers. Complete NLO corrections to $W^\pm Z$, $ZZ$ and $W^+W^-$ scattering have been presented more recently~\cite{Denner:2019tmn,Denner:2020zit,Denner:2021hsa,Denner:2022pwc}. 
While earlier studies of VBS processes have focused on fully leptonic final states, the semileptonic final state has been considered more recently for experimental searches~\cite{ATLAS:2019thr,CMS:2019qfk,CMS:2021qzz} and theoretical calculations~\cite{Denner:2024xul} at LO and beyond~\cite{Jager:2024sij}.

To separate the VBS signal from omni-present background processes, typically cuts on the invariant mass and rapidity difference of the \textit{tagging} jets are employed. 
Signal-to-background ratios can be further improved by restricting the radiation of additional jets in the central rapidity region. A quantitative understanding of non-tagging jets is essential to fully exploit these techniques. 

In this work, we provide a calculation of the production of two on-shell $W^+$ bosons and three jets in the VBS approximation at NLO in QCD and 
matched to a PS. This allows for an improved description of the non-tagging jets in comparison with previous work on same-sign VBS where a third jet could result only from real-emission or parton-shower corrections. 
Furthermore, in our approach the hard region of the fourth jet becomes accessible, since it is generated with real-emission matrix elements.

This article is structured as follows. In Sec.~\ref{sec:process}, we discuss the $pp \to W^+W^+jjj$ process and the approximations used within this work. 
Section~\ref{sec:implementation} provides details of our code implementation. 
In Sec.~\ref{sec:results}, we present results of numerical studies at fixed-order, compare them to PS-matched results, and discuss the impact of PS effects on NLO observables for different shower settings.
We summarize our findings in Sec.~\ref{sec:conclusion}.

\section{Details of the calculation}

%%%%%%%%%%%%%%%%%%%%%%%%%%%%%%%%%%%%%%%%%%%
\subsection{Definition of the process}\label{sec:process}
%%%%%%%%%%%%%%%%%%%%%%%%%%%%%%%%%%%%%%%%%%%
%
In this work, we specifically consider EW $\wpwpjjj$ production in proton-proton collisions at LO (i.e.\ $\order{\alphas\alpha^4}$), and the corresponding NLO-QCD corrections. Strong $\wpwpjjj$ production (with LO contributions of $\order{\alphas^3\alpha^2}$) is not 
considered here. 
Our calculation is based on the VBS approximation, which in the past has been  employed in precision calculations for a vast selection of VBS processes~\cite{Figy:2003nv,Berger:2004pca,Oleari:2003tc,Jager:2006zc,Jager:2006cp,Bozzi:2007ur,Jager:2009xx,Figy:2007kv,Jager:2010aj,Arnold:2010dx,Denner:2012dz,Cacciari:2015jma,Dreyer:2018qbw,Cruz-Martinez:2018rod,cruz-martinezNexttoNexttoLeadingOrderQCD2019}. 
In this framework, it is essentially assumed that VBS processes can be considered as a combination of two independent deep-inelastic scattering processes that do not exchange colour information. Thus, contributions from diagrams with EW boson exchange in the $s$-channel are excluded, as well as interferences of $t$-channel with $u$-channel diagrams. 
The validity of the VBS approximation has been demonstrated explicitly, for instance in Ref.~\cite{Ciccolini:2007ec}, for the related case of $Hjj$ production via VBF at NLO in QCD, for a setup typical for VBS analyses  
in which the two tagging jets are required to exhibit a large separation in rapidity and invariant mass. 
It has also been shown that colour-exchange effects occurring in a full calculation for a $Hjj$ final state at one loop are very strongly suppressed, and can thus safely be neglected~\cite{Andersen:2007mp,Bredenstein:2008tm}. Non-factorizable QCD corrections not accounted for within the VBS approximation have also been found to be small \cite{Dreyer:2020urf,Long:2023mvc,Asteriadis:2023nyl}, and the merging of EW Higgs production processes with different jet multiplicities was considered in Ref.~\cite{Chen:2021phj} resorting to the full NLO-QCD calculation of the EW $Hjjj$ production process~\cite{Campanario:2013fsa}. 
The VBS approximation can be extended to the EW production of a gauge-boson system in association with three jets following the strategy of Ref.~\cite{Figy:2007kv} for VBF-induced $Hjjj$ production. 

%%%%%%%%%%%
For the calculation of the EW $\wpwpjjj$ production process within the VBS approximation we proceed as follows: 
At LO, partonic subprocesses initiated by two light quarks $q_i$,  or a quark and a gluon $g$, such as  
\begin{align}\label{eq:lo_subprocs}
	q_1q_2 \to W^+W^+ q_3q_4 g \quad \text{and} \quad 
	gq_2 \to W^+W^+ q_3q_4\bar q_1 \;, 
\end{align}
and crossing-related processes contribute. Only combinations of quark flavours compatible with the conservation of electric charge are allowed.

\begin{figure}[tbp]
	\centering
	\hbox{\centering
	\subfigure[\label{fig:born_qq_0res_novbs}]{
	\begin{tikzpicture}	
		\begin{feynman}
			\vertex (q1) {\(u\)}; %{\(q_1\)};
			\vertex [below=2cm of q1] (q2) {\(c\)}; %{\(q_2\)};
			\vertex [above right=0.5cm and 4cm of q1] (q3) {\(d\)};%{\(q_3\)};
			\vertex [below right=0.5cm and 4cm of q2] (q4) {\(s\)};%{\(q_4\)};
			\vertex [above right=1.5cm   and 4cm of q1] (g)  {\(g\)};
			\vertex [below right=0.5cm and 4cm of q1] (w1) {\(W^+\)};
			\vertex [above right=0.5cm and 4cm of q2] (w2) {\(W^+\)};
			\vertex [right=1.5cm of q1] (g1);
			\vertex [right=2cm of q1] (wu);
			\vertex [below=2cm of wu] (wd);
			\vertex [above right=0.26cm and 1cm of wu] (wqu);
			\vertex [below right=0.26cm and 1cm of wd] (wqd);
			
			\diagram* {
				(q1) -- [fermion] (wu),
				(wu) -- [fermion] (q3),
				(q2) -- [fermion] (wd),
				(q4) -- [fermion] (q4),
				(g1) -- [gluon] (g),
				(wu) -- [boson, edge label'=\(Z/\gamma\)] (wd),
				(wd) -- [fermion] (q4),
				(wqu) -- [boson] (w1),
				(wqd) -- [boson] (w2),			
			};
		\end{feynman}	
	\end{tikzpicture}
	}
	\subfigure[\label{fig:born_qq_0res_vbs}]{
	\begin{tikzpicture}
		\begin{feynman}
			\vertex (q1) {\(u\)}; %{\(q_1\)};
			\vertex [below=2cm of q1] (q2) {\(c\)}; %{\(q_2\)};
			\vertex [above right=0.5cm and 4cm of q1] (q3) {\(d\)};%{\(q_3\)};
			\vertex [below right=0.5cm and 4cm of q2] (q4) {\(s\)};%{\(q_4\)};
			\vertex [above right=1.5cm   and 4cm of q1] (g)  {\(g\)};
			\vertex [below right=0.5cm and 4cm of q1] (w1) {\(W^+\)};
			\vertex [above right=0.5cm and 4cm of q2] (w2) {\(W^+\)};
			\vertex [right=1.5cm of q1] (g1);
			\vertex [right=2cm of q1] (wu);
			\vertex [below=2cm of wu] (wd);
			\vertex [below=1cm of wu] (wc);
			
			\diagram* {
				(q1) -- [fermion] (wu),
				(wu) -- [fermion] (q3),
				(q2) -- [fermion] (wd),
				(q4) -- [fermion] (q4),
				(g1) -- [gluon] (g),
				(wu) -- [boson] (wd),
				(wd) -- [fermion] (q4),
				(wc) -- [boson] (w1),
				(wc) -- [boson] (w2),			
			};
		\end{feynman}
	\end{tikzpicture}
	}
	\subfigure[\label{fig:born_gq_0res}]{
	\begin{tikzpicture}	
		\begin{feynman}
			\vertex (g) {\(g\)};
			\vertex [below=4cm of g ] (q1) {\(u\)};
			\vertex [right=4cm of g ] (q2) {\(\bar{c}\)};
			\vertex [below=1cm of q2] (w1) {\(W^+\)};
			\vertex [below=1cm of w1] (q3) {\(s\)};
			\vertex [below=1cm of q3] (q4) {\(d\)};
			\vertex [right=4cm of q1] (w2)  {\(W^+\)};
			\vertex [below right=1cm and 1cm of g] (gcc);
			\vertex [below=1cm of gcc] (zcc);
			\vertex [right=1cm of zcc] (csw);
			\vertex [below=2cm of gcc] (uzu);
			\vertex [right=1cm of uzu] (udw);
			
			\diagram* {
				(g) --   [gluon] (gcc),
				(gcc) -- [anti fermion] (q2),
				(gcc) -- [fermion] (zcc),
				(zcc) -- [fermion] (csw),
				(csw) -- [boson] (w1),				
				(csw) -- [fermion] (q3),
				(zcc) -- [boson, edge label'=\(Z\)] (uzu),
				(uzu) -- [anti fermion] (q1),
				(uzu) -- [fermion] (udw),
				(udw) -- [fermion] (q4),
				(udw) -- [boson] (w2),
			};
		\end{feynman}	
	\end{tikzpicture}
	}
	}
	\vspace{0.5cm}
	\hbox{
	% LOWER ROW
	\subfigure[\label{fig:born_qq_1res}]{
	% qq~, s-channel
	\begin{tikzpicture}
		\begin{feynman}
			\vertex (q1)  {\(u\)}; %{\(q_1\)};
			\vertex [below=4cm of q1] (q2) {\(\bar{d}\)};% {\(q_2\)};
			\vertex [below right=2.0cm and 1.0cm of q1] (qqv) ;
			\vertex [right=1.0cm of qqv] (vqq) ;
			
			\vertex [right=4cm of q1] (q3) {\(s\)} ; %{\(q_3\)} ;
			\vertex [right=4cm of q2] (q4) {\(\bar{c}\)} ;%{\(q_4\)} ;
			
			\vertex [below right=1.0cm and 0.5cm of q1] (qg) ;
			\vertex [right=2.0cm of q1] (g) {\(g\)} ;

			\vertex [above right=1.0cm and 1.0cm of vqq ] (qqw1);
			\vertex [below right=1.0cm and 1.0cm of vqq ] (qqw2);
			
			\vertex [below=1.5cm of q3] (w1) {\(W^+\)};
			\vertex [above=1.5cm of q4] (w2) {\(W^+\)};			
			
			\diagram* {
				(q1)   -- [plain]        (qg),
				(qg)   -- [fermion]      (qqv),				
				(q2)   -- [anti fermion] (qqv),
				
				(qqv)  -- [boson] (vqq),

				(vqq)  -- [plain]      (qqw1),
				(vqq)  -- [anti fermion] (qqw2),
				
				(qqw1) -- [fermion] (q3),
				(qqw2) -- [plain] (q4),
				
				(qqw1) -- [boson] (w1),
				(qqw2) -- [boson] (w2),
				
				(qg)   -- [gluon] (g)
			};
		\end{feynman}
	\end{tikzpicture}
	}
	\subfigure[\label{fig:born_qq_3res}]{
	% qq~, three resonances
	\begin{tikzpicture}	
		\begin{feynman}
			\vertex (q1) {\(\bar{d}\)};
			\vertex [below=4cm of q1] (q2) {\(u\)};
			\vertex [right=4cm of q1] (q3) {\(\bar{c}\)};
			\vertex [below=1cm of q3] (q4) {\(s\)};
			\vertex [right=4cm of q2] (g)  {\(g\)};
			\vertex [below=1.8cm of q3] (w1) {\(W^+\)};
			\vertex [below=2.8cm of q3] (w2) {\(W^+\)};
			\vertex [below right=1cm and 1cm of q1] (duw);
			\vertex [above right=1cm and 1cm of q2] (uug);
			\vertex [right=0.8cm of duw] (wzw);
			\vertex [above right=0.2cm and 0.8cm of wzw] (zww);
			\vertex [above right=0.2cm and 0.6cm of zww] (wcs);
			
			\diagram* {
				(q1) --  [anti fermion] (duw),
				(duw) -- [anti fermion] (uug),
				(uug) -- [anti fermion] (q2),
				(uug) -- [gluon] (g),
				
				(duw) -- [boson, edge label=\(W\)] (wzw),
				(wzw) -- [boson, edge label=\(Z\)] (zww),
				(zww) -- [boson, edge label=\(W\)] (wcs),	
				(wzw) -- [boson] (w2),
				(zww) -- [boson] (w1),	
				
				(wcs) -- [anti fermion] (q3),
				(wcs) -- [fermion] (q4),
			};
		\end{feynman}	
	\end{tikzpicture}
	}
	% gq, one resonance
	\subfigure[\label{fig:born_gq_1res}]{
	\begin{tikzpicture}	
		\begin{feynman}
			\vertex (g) {\(g\)};
			\vertex [below=4cm of g ] (q1) {\(u\)};
			\vertex [right=4cm of g ] (q2) {\(d\)};
			\vertex [below=1cm of q2] (w1) {\(W^+\)};
			\vertex [below=1cm of w1] (q3) {\(\bar{c}\)};
			\vertex [below=1cm of q3] (q4) {\(s\)};
			\vertex [right=4cm of q1] (w2)  {\(W^+\)};
			\vertex [below right=1cm and 1cm of g] (gud);
			\vertex [right=1cm of gud] (udw);
			\vertex [above right=1cm and 1cm of q1] (uuz);
			\vertex [right=1cm of uuz] (zcc);
			\vertex [below right=.5cm and .8cm of zcc] (csw);
			
			\diagram* {
				(g) --   [gluon] (gud),
				(gud) -- [fermion] (udw),
				(udw) -- [fermion] (q2),
				(udw) -- [boson] (w1),
				(gud) -- [anti fermion] (uuz),
				(uuz) -- [anti fermion] (q1),
				(uuz) -- [boson, edge label'=\(Z\)] (zcc),
				(zcc) -- [anti fermion] (q3),
				(zcc) -- [anti fermion] (csw),
				(csw) -- [anti fermion] (q4),
				(csw) -- [boson] (w2),
			};
		\end{feynman}	
	\end{tikzpicture}
	}
	}
	\caption{Representative Feynman diagrams for several partonic subprocesses contributing to the EW reaction $pp \to \wpwpjjj$ at LO in QCD. 
}
	\label{fig:FD_born}
\end{figure}
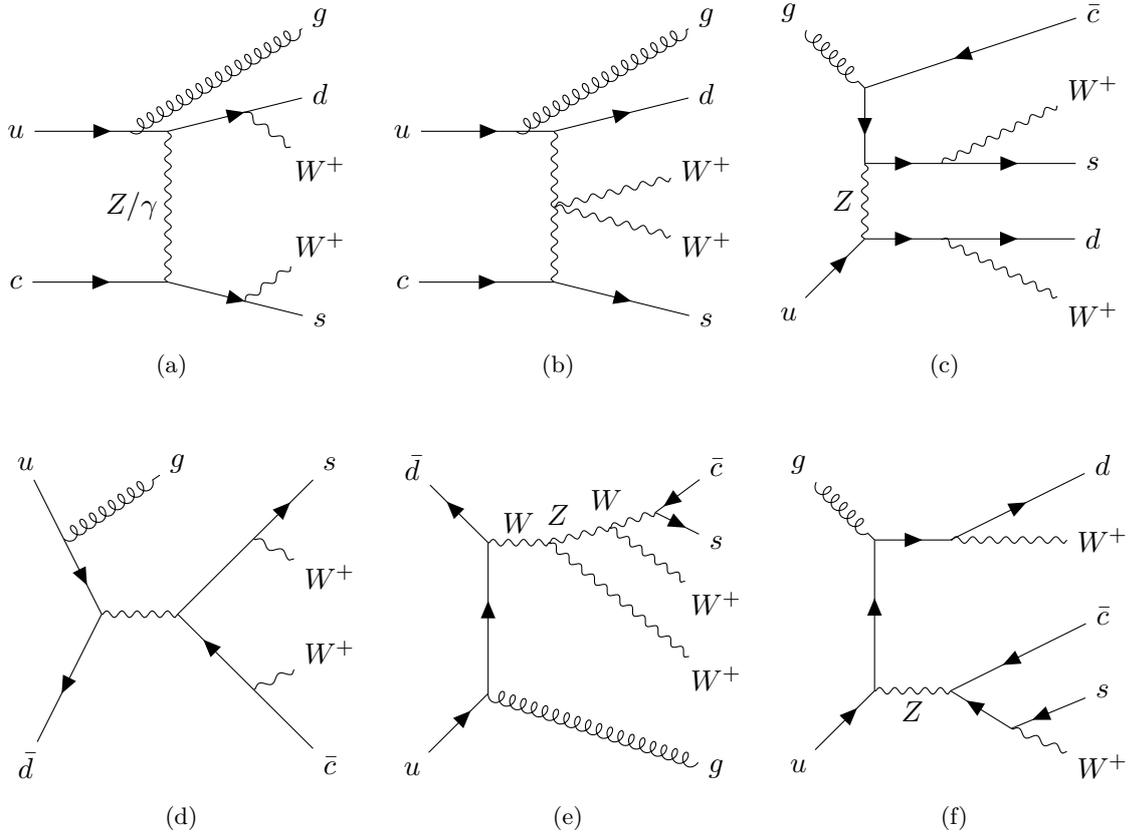

According to the VBS approximation, we only consider Feynman diagrams in which the two initial-state partons are connected by colour-neutral propagators. 
Some representative diagrams that we retain are shown in 
Fig.~(\ref{fig:born_qq_0res_novbs}--\ref{fig:born_gq_0res}), 
while diagrams like those shown in Fig.~(\ref{fig:born_qq_1res}--\ref{fig:born_gq_1res}) are disregarded. 
For  subprocesses with two (anti-)quarks in the initial state, the gluon can be emitted by either fermion line, giving rise to two color structures. 
For  subprocesses with an initial-state gluon, only those diagrams contribute where the incoming gluon splits into a quark-antiquark pair that is connected to the initial-state quark via an EW propagator. For such subprocesses only one color structure occurs. 

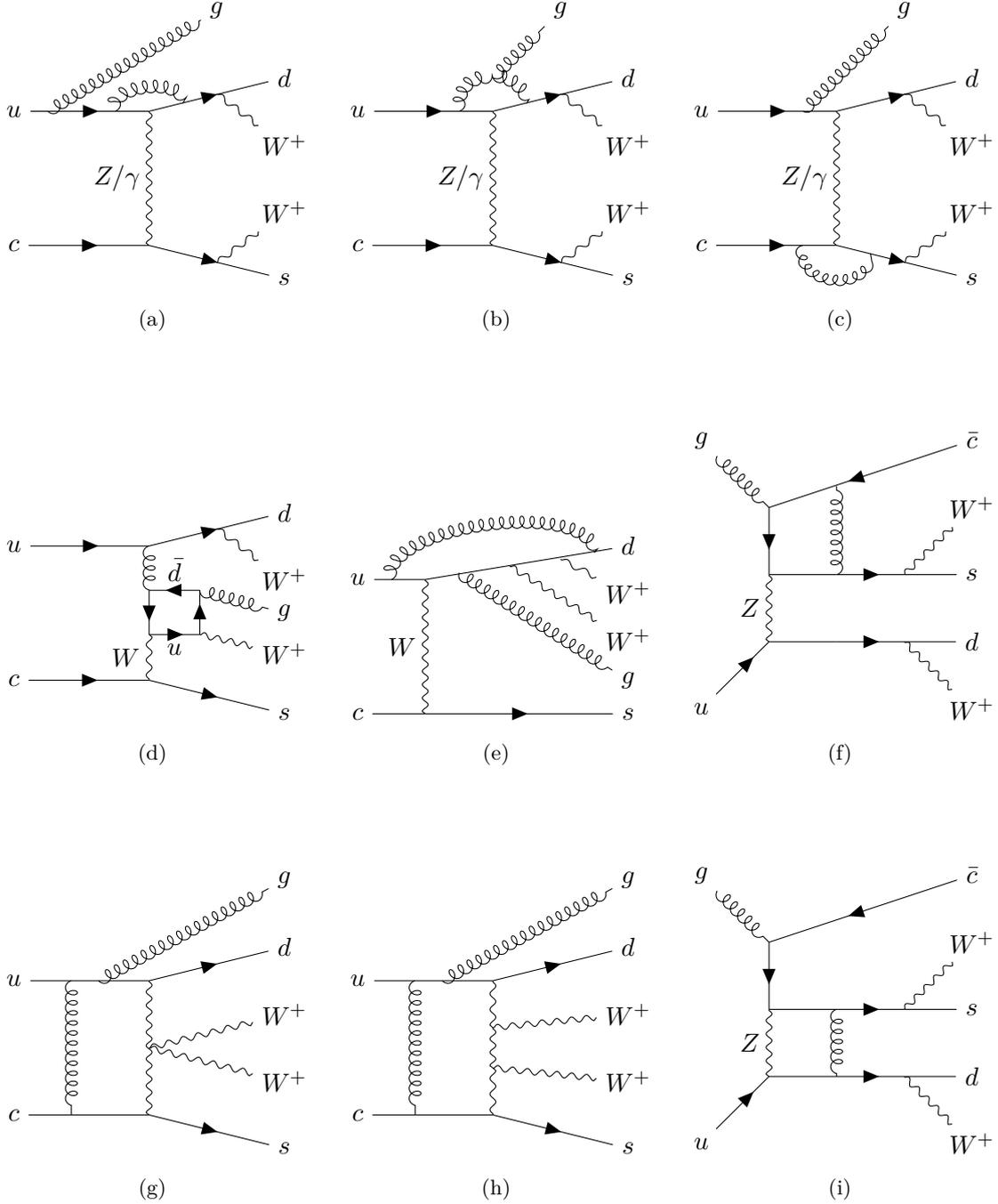
\begin{figure}%[tbp]
	\centering
	\hbox{
		\subfigure[\label{fig:virt_qq_uppervertex_1}]{
		% Triangle vertex correction
		\begin{tikzpicture}	
			\begin{feynman}
				\vertex (q1) {\(u\)};
				\vertex [below=2cm of q1] (q2) {\(c\)};
				\vertex [above right=0.5cm and 4cm of q1] (q3) {\(d\)};
				\vertex [below right=0.5cm and 4cm of q2] (q4) {\(s\)};
				\vertex [above right=1.5cm and 3cm of q1] (g)  {\(g\)};
				\vertex [below right=0.5cm and 4cm of q1] (w1) {\(W^+\)};
				\vertex [above right=0.5cm and 4cm of q2] (w2) {\(W^+\)};
				\vertex [right=0.5cm of q1] (g1);
				\vertex [right=2cm of q1] (wu);
				\vertex [below=2cm of wu] (wd);
				\vertex [left=0.5cm of wu] (q1l);
				\vertex [above right=0.12cm and 0.5cm of wu] (q3l);
				\vertex [above right=0.26cm and 1cm of wu] (wqu);
				\vertex [below right=0.26cm and 1cm of wd] (wqd);
				
				\diagram* {
					(q1) -- [fermion] (wu),
					(wu) -- [fermion] (q3),
					(q2) -- [fermion] (wd),
					(q1l) -- [gluon, half left,looseness=1] (q3l),
					(g1) -- [gluon] (g),
					(wu) -- [boson, edge label'=\(Z/\gamma\)] (wd),
					(wd) -- [fermion] (q4),
					(wqu) -- [boson] (w1),
					(wqd) -- [boson] (w2),			
				};
			\end{feynman}
		\end{tikzpicture}
		}
		%
		% Box vertex correction, triple gluon coupling
		\subfigure[\label{fig:virt_qq_uppervertex_2}]{
		\begin{tikzpicture}
			\begin{feynman}
				\vertex (q1) {\(u\)};
				\vertex [below=2cm of q1] (q2) {\(c\)};
				\vertex [above right=0.5cm and 4cm of q1] (q3) {\(d\)};
				\vertex [below right=0.5cm and 4cm of q2] (q4) {\(s\)};
				\vertex [above right=1.5cm and 3cm of q1] (g)  {\(g\)};
				\vertex [below right=0.5cm and 4cm of q1] (w1) {\(W^+\)};
				\vertex [above right=0.5cm and 4cm of q2] (w2) {\(W^+\)};
				\vertex [above right=0.5cm and 2cm of q1] (ggg);
				
				\diagram* {
					(q1) -- [fermion] (wu),
					(wu) -- [fermion] (q3),
					(q2) -- [fermion] (wd),
					(q1l) -- [gluon, quarter left,looseness=1.2] (ggg),
					(ggg) -- [gluon, quarter left,looseness=0.9] (q3l),		
					(ggg) -- [gluon] (g),
					(wu) -- [boson, edge label'=\(Z/\gamma\)] (wd),
					(wd) -- [fermion] (q4),
					(wqu) -- [boson] (w1),
					(wqd) -- [boson] (w2),			
				};
			\end{feynman}	
		\end{tikzpicture}
		}
%		%
%		% Pentagon, upper line
%		\begin{tikzpicture}	
%			\begin{feynman}
%				\vertex (q1) {\(q_1\)};
%				\vertex [below=2cm of q1] (q2) {\(q_2\)};
%				\vertex [above right=0.5cm and 4cm of q1] (q3) {\(q_3\)};
%				\vertex [below right=0.5cm and 4cm of q2] (q4) {\(q_4\)};
%				\vertex [below right=1cm and 4cm of q1] (g)  {\(g\)};
%				\vertex [below right=0.3cm and 4cm of q1] (w1) {\(W^+\)};
%				\vertex [above right=0.5cm and 4cm of q2] (w2) {\(W^+\)};
%				\vertex [above right=0.12cm and 0.5cm of wu] (g1);
%				\vertex [right=2cm of q1] (wu);
%				\vertex [below=2cm of wu] (wd);
%				\vertex [left=0.5cm of wu] (q1l);
%				\vertex [above right=0.32cm and 1.5cm of wu] (q3l);
%				\vertex [above right=0.26cm and 1cm of wu] (wqu);
%				\vertex [below right=0.26cm and 1cm of wd] (wqd);
%				
%				\diagram* {
%					(q1) -- [fermion] (wu),
%					(wu) -- [fermion] (q3),
%					(q2) -- [fermion] (wd),
%					(q1l) -- [gluon, half left,looseness=0.7] (q3l),
%					(g1) -- [gluon] (g),
%					(wu) -- [boson, edge label'=\(Z/\gamma\)] (wd),
%					(wd) -- [fermion] (q4),
%					(wqu) -- [boson] (w1),
%					(wqd) -- [boson] (w2),			
%				};
%			\end{feynman}
%		\end{tikzpicture}
				%
		%
		% lower line
		\subfigure[\label{fig:virt_qq_lowervertex_1}]{
		\begin{tikzpicture}	
			\begin{feynman}
				\vertex (q1) {\(u\)};
				\vertex [below=2cm of q1] (q2) {\(c\)};
				\vertex [above right=0.5cm and 4cm of q1] (q3) {\(d\)};
				\vertex [below right=0.5cm and 4cm of q2] (q4) {\(s\)};
				\vertex [above right=1.5cm and 3cm of q1] (g)  {\(g\)};
				\vertex [below right=0.5cm and 4cm of q1] (w1) {\(W^+\)};
				\vertex [above right=0.5cm and 4cm of q2] (w2) {\(W^+\)};
				\vertex [right=1.5cm of q1] (g1);
				\vertex [right=2cm of q1] (wu);
				\vertex [below=2cm of wu] (wd);
				\vertex [left=0.5cm of wd] (q2l);
				\vertex [below right=0.12cm and 0.5cm of wd] (q4l);
				\vertex [above right=0.26cm and 1cm of wu] (wqu);
				\vertex [below right=0.26cm and 1cm of wd] (wqd);
				
				\diagram* {
					(q1) -- [fermion] (wu),
					(wu) -- [fermion] (q3),
					(q2) -- [fermion] (wd),
					%			(q4) -- [fermion] (q4),
					(q2l) -- [gluon, half right] (q4l),
					(g1) -- [gluon] (g),
					(wu) -- [boson, edge label'=\(Z/\gamma\)] (wd),
					(wd) -- [fermion] (q4),
					(wqu) -- [boson] (w1),
					(wqd) -- [boson] (w2),			
				};
			\end{feynman}	
		\end{tikzpicture}
	}
	}
	\vspace{1cm}
	\hbox{
%		%
%		% Triangle (internal quark loop)
%		\begin{tikzpicture}
%			\begin{feynman}
%				\vertex (q1) {\(q_1\)};
%				\vertex [below=2cm of q1] (q2) {\(q_2\)};
%				\vertex [above right=0.5cm and 4cm of q1] (q3) {\(q_3\)};
%				\vertex [below right=0.5cm and 4cm of q2] (q4) {\(q_4\)};
%				\vertex [below right=1cm and   4cm of q1] (g)  {\(g\)};
%				\vertex [below right=0.5cm and 4cm of q1] (w1) {\(W^+\)};
%				\vertex [above right=0.5cm and 4cm of q2] (w2) {\(W^+\)};
%				%		
%				\vertex [right=2cm of q1] (q1gq3);
%				\vertex [below=0.66cm of wu] (lu);
%				\vertex [below=0.66cm of lu] (ld);
%				\vertex [below=2cm of q1gq3] (q2zq4);
%				\vertex [below right=1cm and 0.75cm of q1gq3] (qqg);
%				
%				\diagram* {
%					(q1)    -- [fermion] (q1gq3),
%					(q1gq3) -- [fermion] (q3),
%					(q2)    -- [fermion] (q2zq4),
%					(q2zq4) -- [fermion] (q4),
%					
%					(wqu)   -- [boson] (w1),
%					(wqd)   -- [boson] (w2),			
%					
%					(q1gq3) -- [gluon] (lu),
%					(q2zq4) -- [boson] (ld),
%					
%					(lu)  -- [fermion] (ld),
%					(ld)  -- [fermion] (qqg),
%					(qqg) -- [fermion] (lu),
%					
%					(qqg) -- [gluon] (g)		
%					
%				};
%			\end{feynman}	
%		\end{tikzpicture}
		%
		% Box (internal quark loop)
		\subfigure[\label{fig:virt_qq_quarkloop}]{
		\begin{tikzpicture}
			\begin{feynman}
				\vertex (q1) {\(u\)};
				\vertex [below=2cm of q1] (q2) {\(c\)};
				\vertex [above right=0.5cm and 4cm of q1] (q3) {\(d\)};
				\vertex [below right=0.5cm and 4cm of q2] (q4) {\(s\)};
				\vertex [below right=1cm and   4cm of q1] (g)  {\(g\)};
				\vertex [below right=0.5cm and 4cm of q1] (w1) {\(W^+\)};
				\vertex [above right=0.4cm and 4cm of q2] (w2) {\(W^+\)};
				\vertex [right=2cm of q1] (q1gq3);
				\vertex [below=0.66cm of wu] (lu);
				\vertex [below=0.66cm of lu] (ld);
				\vertex [below=2cm of q1gq3] (q2zq4);
				
				\vertex [right=0.75cm of lu] (qqg);
				\vertex [right=0.75cm of ld] (qqw);
				
				\diagram*{
					(q1)    -- [fermion] (q1gq3),
					(q1gq3) -- [fermion] (q3),
					(q2)    -- [fermion] (q2zq4),
					(q2zq4) -- [fermion] (q4),
					
					(wqu)   -- [boson] (w1),
					(qqw)   -- [boson] (w2),			
					
					(q1gq3) -- [gluon] (lu),
					(q2zq4) -- [boson, edge label=\(W\)] (ld),
					
					(lu)  -- [fermion] (ld),
					(ld)  -- [fermion, edge label'=\(u\)] (qqw),
					(qqw) -- [fermion] (qqg),
					(qqg) -- [fermion, edge label'=\(\bar{d}\)] (lu),
					
					(qqg) -- [gluon] (g),
				};
			\end{feynman}	
		\end{tikzpicture}
		}
		%
		% Hexagons
		\subfigure[\label{fig:virt_qq_hexagon}]{
		\begin{tikzpicture}	
			\begin{feynman}
				\vertex (q1) {\(u\)};
				\vertex [below=2cm of q1] (q2) {\(c\)};
				\vertex [above right=0.5cm and 4cm of q1] (q3) {\(d\)};
				\vertex [right=4cm of q2] (q4) {\(s\)};
				\vertex [below=0.7cm of q3] (w1) {\(W^+\)};
				\vertex [below=1.4cm of q3] (w2) {\(W^+\)};
				\vertex [below=2.0cm of q3] (g)  {\(g\)};
				\vertex [right=1cm of q1] (wu);
				\vertex [below=2cm of wu] (wd);
				\vertex [right=0.5cm of q1] (q1l);
				\vertex [below left=0.1cm and 0.5cm of q3] (q3l);
				\vertex [below left=0.2cm and 1.0cm of q3] (wqu);
				\vertex [below left=0.3cm and 1.75cm of q3] (wqd);
				\vertex [below left=0.4cm and 2.5cm of q3] (g1);
				
				\diagram* {
					(q1) -- [plain] (wu),
					(wu) -- [plain] (q3),
					(q2) -- [plain] (wd),
					(q1l) -- [gluon, half left,looseness=0.7] (q3l),
					(g1) -- [gluon] (g),
					(wu) -- [boson, edge label'=\(W\)] (wd),
					(wd) -- [fermion] (q4),
					(wqu) -- [boson] (w1),
					(wqd) -- [boson] (w2),			
				};
			\end{feynman}
		\end{tikzpicture}
		}
		\subfigure[\label{fig:virt_gq_box}]{
		% gq virtual no gluon exchange
		\begin{tikzpicture}	
			\begin{feynman}
				\vertex (g) {\(g\)};
				\vertex [below=4cm of g ] (q1) {\(u\)};
				\vertex [right=4cm of g ] (q2) {\(\bar{c}\)};
				\vertex [below=1cm of q2] (w1) {\(W^+\)};
				\vertex [below=1cm of w1] (q3) {\(s\)};
				\vertex [below=1cm of q3] (q4) {\(d\)};
				\vertex [right=4cm of q1] (w2)  {\(W^+\)};			
				\vertex [below right=1cm and 1cm of g] (gcc);
				\vertex [below=1cm of gcc] (zcc);
				\vertex [right=2cm of zcc] (csw);
				\vertex [below=2cm of gcc] (uzu);
				\vertex [right=2cm of uzu] (udw);
				
				\vertex [right=1cm of zcc] (csg);
				\vertex [right=1cm of uzu] (udg);
				
				\vertex [above right=0.32cm and 1cm of gcc] (gccl);
				
				\diagram* {
					(g) --   [gluon] (gcc),
					(gcc) -- [anti fermion] (q2),
					(gcc) -- [fermion] (zcc),
					(zcc) -- [plain] (csg),
					
					(csg) -- [fermion] (csw),
					(csg) -- [gluon] (gccl),
					
					(csw) -- [boson] (w1),				
					(csw) -- [plain] (q3),
					(zcc) -- [boson, edge label'=\(Z\)] (uzu),
					(uzu) -- [anti fermion] (q1),
					(uzu) -- [plain] (udg),
					(udg) -- [fermion] (udw),
					(udw) -- [plain] (q4),
					(udw) -- [boson]   (w2),
				};
			\end{feynman}	
		\end{tikzpicture}
		}
	}
	\vspace{1cm}
	\hbox{	
%		%
%		% Pentagon
%		\begin{tikzpicture}	
%			\begin{feynman}
%				\vertex (q1) {\(q_1\)};
%				\vertex [right=.85cm of q1] (glu);
%				\vertex [right=1.25cm of q1] (gex);
%				\vertex [right=2.0cm of q1] (wu);
%				\vertex [above right=0.5cm and 4cm of q1] (q3) {\(q_3\)};
%				\vertex [above right=0.26cm and 1cm of wu] (wqu);		
%				\vertex [below right=0.5cm and 4cm of q1] (w1) {\(W^+\)};
%				\vertex [above right=1.5cm   and 4cm of q1] (g)  {\(g\)};		
%			
%				\vertex [below=2cm of q1] (q2) {\(q_2\)};
%				\vertex [below=2cm of glu] (gld);
%				\vertex [below=2cm of wu] (wd);
%				\vertex [below right=0.26cm and 1cm of wd] (wqd);
%				\vertex [above right=0.5cm and 4cm of q2] (w2) {\(W^+\)};
%				\vertex [below right=0.5cm and 4cm of q2] (q4) {\(q_4\)};
%			
%			
%				\diagram* {
%					(q1)  -- [plain] (glu),
%					(glu) -- [plain] (gex),
%					(gex) -- [plain]  (wu),
%					(wu)  -- [fermion]  (wqu),
%					(wqu) -- [plain] (q3),
%					(q2)  -- [plain] (wd),
%					(q2)  -- [plain] (gld),
%					(gld) -- [plain] (wd),
%					(q4)  -- [fermion] (q4),
%					(gex) -- [gluon] (g),
%					(wu)  -- [boson, edge label=\(Z/\gamma\)] (wd),
%					(wd)  -- [fermion] (wqd),
%					(wqd) -- [plain] (q4),
%					(wqu) -- [boson] (w1),
%					(wqd) -- [boson] (w2),		
%					(glu) -- [gluon] (gld),	
%				};
%			\end{feynman}	
%		\end{tikzpicture}
		%
		% Hexa
		\subfigure[\label{fig:virt_qq_hexagon_gluonex}]{
		\begin{tikzpicture}
			\begin{feynman}
				\vertex (q1) {\(u\)};
				\vertex [right=.85cm of q1] (glu);
				\vertex [right=1.25cm of q1] (gex);
				\vertex [below=2cm of q1] (q2) {\(c\)};
				\vertex [below=2cm of glu] (gld);
				\vertex [above right=0.5cm and 4cm of q1] (q3) {\(d\)};
				\vertex [below right=0.5cm and 4cm of q2] (q4) {\(s\)};
				\vertex [above right=1.5cm   and 4cm of q1] (g)  {\(g\)};
				\vertex [below right=0.5cm and 4cm of q1] (w1) {\(W^+\)};
				\vertex [above right=0.5cm and 4cm of q2] (w2) {\(W^+\)};
				\vertex [right=2cm of q1] (wu);
				\vertex [below=2cm of wu] (wd);
				\vertex [below=1cm of wu] (wc);
			
				\diagram* {
					(q1)  -- [plain] (glu),
					(glu) -- [plain] (gex),
					(gex) -- [plain]  (wu),
					(wu)  -- [fermion] (q3),
					(q2)  -- [plain] (gld),
					(gld) -- [plain] (wd),
					(q4)  -- [fermion] (q4),
					(gex) -- [gluon] (g),
					(wu)  -- [boson] (wd),
					(wd)  -- [fermion] (q4),
					(wc)  -- [boson] (w1),
					(wc)  -- [boson] (w2),
					(glu) -- [gluon] (gld),			
				};
			\end{feynman}
		\end{tikzpicture}
		}
		%
		% Septa
		\subfigure[\label{fig:virt_qq_heptagon_gluonex}]{
		\begin{tikzpicture}	
			\begin{feynman}
				\vertex (q1) {\(u\)};
				\vertex [right=.85cm of q1] (glu);
				\vertex [right=1.25cm of q1] (gex);
				\vertex [below=2cm of q1] (q2) {\(c\)};
				\vertex [below=2cm of glu] (gld);
				\vertex [above right=0.5cm and 4cm of q1] (q3) {\(d\)};
				\vertex [below right=0.5cm and 4cm of q2] (q4) {\(s\)};
				\vertex [above right=1.5cm   and 4cm of q1] (g)  {\(g\)};
				\vertex [below right=0.5cm and 4cm of q1] (w1) {\(W^+\)};
				\vertex [above right=0.5cm and 4cm of q2] (w2) {\(W^+\)};
				\vertex [right=2cm of q1] (wu);
				\vertex [below=2cm of wu] (wd);
				\vertex [below=0.70cm of wu] (whu);
				\vertex [above=0.70cm of wd] (whd);
			
				\diagram* {
					(q1)  -- [plain]   (glu),
					(glu) -- [plain]   (gex),
					(gex) -- [plain]   (wu),
					(wu)  -- [fermion] (q3),
					(q2)  -- [plain]   (gld),
					(gld) -- [plain]   (wd),
					(q4)  -- [fermion] (q4),
					(gex) -- [gluon]   (g),
					(wd)  -- [fermion] (q4),
					(wu)  -- [boson]   (whu),
					(wd)  -- [boson]   (whd),		
					(whu) -- [boson]   (w1),
					(whd) -- [boson]   (w2),
				%	(whu) -- [boson, edge label'=\(Z/\gamma\)] (whd),
					(whu) -- [boson]   (whd),
					(glu) -- [gluon]   (gld),			
				};
			\end{feynman}	
		\end{tikzpicture}
		}
		\subfigure[\label{fig:virt_gq_box_gluonex}]{
		% gq virtual loop gluon exchange
		\begin{tikzpicture}	
			\begin{feynman}
				\vertex (g) {\(g\)};
				\vertex [below=4cm of g ] (q1) {\(u\)};
				\vertex [right=4cm of g ] (q2) {\(\bar{c}\)};
				\vertex [below=1cm of q2] (w1) {\(W^+\)};
				\vertex [below=1cm of w1] (q3) {\(s\)};
				\vertex [below=1cm of q3] (q4) {\(d\)};
				\vertex [right=4cm of q1] (w2)  {\(W^+\)};
				\vertex [below right=1cm and 1cm of g] (gcc);
				\vertex [below=1cm of gcc] (zcc);
				\vertex [right=2cm of zcc] (csw);
				\vertex [below=2cm of gcc] (uzu);
				\vertex [right=2cm of uzu] (udw);
				
				\vertex [right=1cm of zcc] (csg);
				\vertex [right=1cm of uzu] (udg);
				
				\diagram* {
					(g) --   [gluon] (gcc),
					(gcc) -- [anti fermion] (q2),
					(gcc) -- [fermion] (zcc),
					(zcc) -- [plain] (csg),
					
					(csg) -- [fermion] (csw),
					(csg) -- [gluon] (udg),
					
					(csw) -- [boson] (w1),				
					(csw) -- [plain] (q3),
					(zcc) -- [boson, edge label'=\(Z\)] (uzu),
					(uzu) -- [anti fermion] (q1),
					(uzu) -- [plain] (udg),
					(udg) -- [fermion] (udw),
					(udw) -- [plain] (q4),
					(udw) -- [boson]   (w2),
				};
			\end{feynman}	
		\end{tikzpicture}
		}
	}
	\caption{\label{fig:FD_virt}
Representative one-loop diagrams for several partonic subprocesses contributing to the EW reaction $pp \to \wpwpjjj$. 	
}
	\end{figure}

%%%%%%%%
Virtual corrections arise from the interference of the LO contributions discussed above with one-loop diagrams. According to the VBS approximation, we only take loop diagrams with no colour flow between external fermion lines into account. Figure~\ref{fig:FD_virt} shows some examples of included (\ref{fig:virt_qq_uppervertex_1}--\ref{fig:virt_gq_box}) and disregarded (\ref{fig:virt_qq_hexagon_gluonex}--\ref{fig:virt_gq_box_gluonex}) diagrams. 
The considered loop diagrams can involve up to six propagators requiring the evaluation of hexagon integrals. 

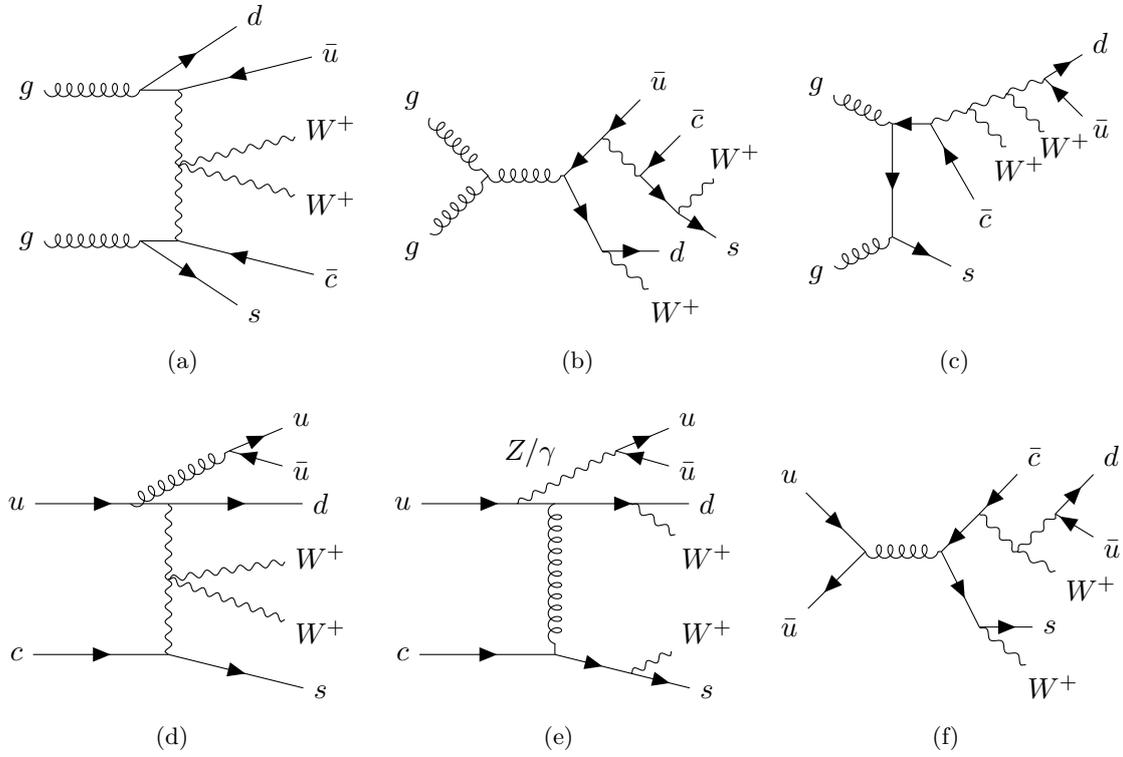
\begin{figure}%[tbp]%[h]
	\centering
% Reals
	% vbs
	\subfigure[\label{fig:real_gg_0res}]{
	\begin{tikzpicture}
		\begin{feynman}
			\vertex (g1) {\(g\)};
			\vertex [below=2cm of g1] (g2) {\(g\)};
			\vertex [above right=1cm and 3cm of g1] (q1)  {\(d\)};
			\vertex [below right=3cm and 3cm of g1] (q2)  {\(s\)};	
			\vertex [above right=0.5cm and 4cm of g1] (q3) {\(\bar{u}\)};
			\vertex [below right=0.5cm and 4cm of g2] (q4) {\(\bar{c}\)};
			\vertex [below right=0.5cm and 4cm of g1] (w1) {\(W^+\)};
			\vertex [above right=0.5cm and 4cm of g2] (w2) {\(W^+\)};
			
			\vertex [right=1.5cm of g1] (gqqu);
			\vertex [right=1.5cm of g2] (gqqd);		
			
			\vertex [right=2cm of g1] (wu);
			\vertex [below=2cm of wu] (wd);
			\vertex [below=1cm of wu] (wc);
			
			\diagram* {
				(g1)   -- [gluon]   (gqqu),
				(gqqu) -- [fermion]   (q1),
				(gqqu) -- [plain]   (wu),
				(wu)   -- [anti fermion] (q3),
				
				(g2)   -- [gluon]   (gqqd),
				(gqqd) -- [fermion]   (q2),
				(gqqd) -- [plain]   (wd),
				(wd)   -- [anti fermion] (q4),
				
				(wu)  -- [boson] (wc),
				(wc)  -- [boson] (wd),
				(wc)  -- [boson] (w1),
				(wc)  -- [boson] (w2), 		
			};
		\end{feynman}
	\end{tikzpicture}
	}
	% ggg-vertex
	\subfigure[\label{fig:real_gg_1res}]{
	\begin{tikzpicture}
		\begin{feynman}
			\vertex (g1) {\(g\)};
			\vertex [below=2cm of g1] (g2)  {\(g\)};
			\vertex [below right=1.0cm and 1.0cm of g1] (ggg) ;
			\vertex [right=1.0cm of ggg] (gqq) ;	
			\vertex [above right=0.5cm and 0.5cm of gqq] (qqz) ;
			\vertex [below right=0.5cm and 0.5cm of qqz] (zqq) ;
			\vertex [below right=0.5cm and 0.5cm of zqq] (qqw1) ;

			\vertex [below right=1.0cm and 0.5cm of gqq] (qqw2) ;
			\vertex [below right=0.5cm and 0.5cm of qqz] (zqq) ;
			\vertex [below right=0.5cm and 0.5cm of zqq] (qqw1) ;
											
			\vertex [above right=0.5cm and 0.5cm of qqz] (q1) {\(\bar{u}\)};
			\vertex [below right=0.5cm and 0.5cm of q1]  (q2) {\(\bar{c}\)};	
			\vertex [below right=0.5cm and 0.5cm of q2]  (w1) {\(W^+\)};
%			\vertex [below right=0.5cm and 0.5cm of qqw1] (q4) {\(q_4\)};
			\vertex [below right=0.25cm and 0.5cm of qqw1] (q4) {\(s\)};

%			\vertex [above right=0.5cm and 0.5cm of qqw2] (q3) {\(q_3\)};
			\vertex [right=0.75cm of qqw2] (q3) {\(d\)};
			\vertex [below right=0.5cm and 0.5cm of qqw2] (w2) {\(W^+\)};

			\diagram* {
				(g1)  -- [gluon]   (ggg),
				(g2)  -- [gluon]   (ggg),
				
				(ggg) -- [gluon]   (gqq),
				
				(gqq) -- [anti fermion] (qqz),
				(qqz) -- [anti fermion] (q1),
				(qqz) -- [boson] (zqq),
				(zqq) -- [anti fermion] (q2),
				(zqq) -- [fermion] (qqw1),
				(qqw1) -- [fermion] (q4),
				(qqw1) -- [boson] (w1),
								
				(gqq)  -- [fermion] (qqw2),
				(qqw2) -- [fermion] (q3),
				(qqw2) -- [boson] (w2),
			};
		\end{feynman}
	\end{tikzpicture}
	}
	%
	% 3 res
	\subfigure[\label{fig:real_gg_3res}]{
	\begin{tikzpicture}	
		\begin{feynman}
			\vertex (g1) {\(g\)};
			\vertex [below right=0.5cm and 1.0cm of g1] (gqqu) ;
			\vertex [right=0.5cm of gqqu] (qwq) ; 
			\vertex [below right=1.0cm and 0.5cm of qwq] (q3) {\(\bar{c}\)} ;
			\vertex [above right=0.2cm and 0.5cm of qwq] (wzw) ;
	%		\vertex [below right=0.5cm and 0.5cm of wzw] (w1) {\(W^+\)} ;
			\vertex [below right=0.5cm and 0.5cm of wzw] (w1) ;
			\vertex [below right=0.5cm and 0.2cm of wzw] (w1lab) {\(W^+\)} ;
			\vertex [above right=0.2cm and 0.5cm of wzw] (zww);
	%		\vertex [below right=0.5cm and 0.5cm of zww] (w2) {\(W^+\)} ;
			\vertex [below right=0.5cm and 0.5cm of zww] (w2) ;
			\vertex [below right=0.4cm and 0.3cm of zww] (w2lab) {\(W^+\)} ;
			\vertex [above right=0.2cm and 0.5cm of zww] (wqq) ;
			\vertex [above right=0.2cm and 0.5cm of wqq] (q1) {\(d\)} ;
	%		\vertex [below right=0.5cm and 0.5cm of wqq] (q2) {\(q_2\)} ;
			\vertex [below right=0.5cm and 0.5cm of wqq] (q2) ;
			\vertex [below right=0.4cm and 0.5cm of wqq] (q2lab) {\(\bar{u}\)} ;
									
			\vertex [below=2.5cm of g1] (g2) {\(g\)} ;
			\vertex [above right=0.5cm and 1.0cm of g2] (gqqd) ;
			\vertex [right=2cm of g2] (q4) {\(s\)} ;
			\vertex [below=1.0cm of gqqd] (x1) {\(\)};
					
			\diagram* {
				(g1)   -- [gluon]   (gqqu),
				(gqqu) -- [anti fermion] (qwq),
				(qwq)  -- [anti fermion] (q3),	
				(qwq)  -- [boson]   (wzw),
				(wzw)  -- [boson]   (zww),
				(wzw)  -- [boson]   (w1),
				(zww)  -- [boson]   (wqq),
				(zww)  -- [boson]   (w2),
				(wqq)  -- [fermion] (q1),
				(wqq)  -- [anti fermion] (q2),
				
				(gqqu) -- [fermion] (gqqd),
				
				(g2)   -- [gluon] (gqqd),
				(gqqd) -- [fermion] (q4),
			};
		\end{feynman}	
	\end{tikzpicture}
	}
	\subfigure[\label{fig:real_qq_0res}]{
	% g > qq~
	\begin{tikzpicture}
		\begin{feynman}
			\vertex (q1) {\(u\)};
			\vertex [below=2cm of q1] (q2) {\(c\)};
%			\vertex [above right=0.5cm and 4cm of q1] (q3) {\(q_3\)};
			\vertex [right=4cm of q1] (q3) {\(d\)};
			\vertex [below right=0.5cm and 4cm of q2] (q4) {\(s\)};
			\vertex [right=1.5cm of q1] (g1);
			\vertex [above right=0.7cm and 1.3cm of g1] (gqq);
%			\vertex [above right=0.3cm and 0.5cm of gqq] (q5) {\(q_5\)};
%			\vertex [below right=0.1cm and 0.5cm of gqq] (q6) {\(q_6\)};
			\vertex [above right=0.3cm and 0.7cm of gqq] (q5) ;
			\vertex [below right=0.2cm and 0.7cm of gqq] (q6) ;
			\vertex [above right=0.2cm and 0.7cm of gqq] (q5lab) {\(u\)};
			\vertex [below right=0.0cm and 0.7cm of gqq] (q6lab) {\(\bar{u}\)};
			\vertex [below right=0.7cm and 4cm of q1] (w1) {\(W^+\)};
			\vertex [above right=0.3cm and 4cm of q2] (w2) {\(W^+\)};
			\vertex [right=2cm of q1] (wu);
			\vertex [below=2cm of wu] (wd);
			\vertex [below=1cm of wu] (wc);
			
			\diagram* {
				(q1) -- [fermion] (wu),
				(wu) -- [fermion] (q3),
				(q2) -- [fermion] (wd),
				(q4) -- [fermion] (q4),
				(g1) -- [gluon] (gqq),
				(wu) -- [boson] (wd),
				(wd) -- [fermion] (q4),
				(wc) -- [boson] (w1),
				(wc) -- [boson] (w2),			
				(gqq) -- [fermion] (q5),
				(gqq) -- [anti fermion] (q6),
			};
		\end{feynman}
	\end{tikzpicture}
	}
	\subfigure[\label{fig:real_qq_1res}]{
	% gamma > qq~
	\begin{tikzpicture}
		\begin{feynman}
			\vertex (q1) {\(u\)};
			\vertex [below=2cm of q1] (q2) {\(c\)};
			\vertex [right=4cm of q1] (q3) {\(d\)};
			\vertex [below right=0.5cm and 4cm of q2] (q4) {\(s\)};
			\vertex [right=1.5cm of q1] (g1);
			\vertex [above right=0.7cm and 1.3cm of g1] (gqq);
			\vertex [above right=0.3cm and 0.7cm of gqq] (q5) ;
			\vertex [below right=0.2cm and 0.7cm of gqq] (q6) ;
			\vertex [above right=0.2cm and 0.7cm of gqq] (q5lab) {\(u\)};
			\vertex [below right=0.0cm and 0.7cm of gqq] (q6lab) {\(\bar{u}\)};
			\vertex [below right=0.7cm and 4cm of q1] (w1) {\(W^+\)};
			\vertex [above right=0.3cm and 4cm of q2] (w2) {\(W^+\)};
			\vertex [right=2cm of q1] (wu);
			\vertex [below=2cm of wu] (wd);
			\vertex [right=1cm of wu] (qwqu);
			\vertex [below right=0.25cm and 1.0cm of wd] (qwqd);
			
			\diagram* {
				(q1) -- [fermion] (wu),
				(wu) -- [fermion] (q3),
				(q2) -- [fermion] (wd),
				(q4) -- [fermion] (q4),
				(g1) -- [boson,edge label=\(Z/\gamma\)] (gqq),
				(wu) -- [gluon] (wd),
%				(wd) -- [fermion] (q4),
				(wd) -- [fermion] (qwqd),
				(qwqd) -- [fermion] (q4),				
				(qwqu) -- [boson] (w1),
				(qwqd) -- [boson] (w2),			
				(gqq) -- [fermion] (q5),
				(gqq) -- [anti fermion] (q6),
			};
		\end{feynman}
	\end{tikzpicture}
	}
	\subfigure[\label{fig:real_qq_2res}]{
	% 2 ew res
	\begin{tikzpicture}
		\begin{feynman}
			\vertex (q1) {\(u\)};
			\vertex [below=2cm of q1] (q2) {\(\bar{u}\)};
			\vertex [below right=1.0cm and 1.0cm of q1]  (qqg) ;
			\vertex [right=1.0cm of qqg] (gqq) ;
			\vertex [above right=0.5cm and 0.5cm of gqq] (qqa) ;

			\vertex [above right=0.5cm and 0.5cm of qqa] (q3) {\(\bar{c}\)} ;
			\vertex [below right=0.5cm and 0.5cm of qqa] (aww) ;
			\vertex [above right=0.5cm and 0.5cm of aww] (wqq) ;
			\vertex [below right=0.2cm and 0.5cm of aww] (w1) {\(W^+\)} ;	
			\vertex [above right=0.5cm and 0.5cm of wqq] (q5) {\(d\)} ;
			\vertex [below right=0.2cm and 0.5cm of wqq] (q6) {\(\bar{u}\)} ;
			
			\vertex [below right=1.0cm and 0.5cm of gqq] (qqw) ;				
%			\vertex [above right=0.0cm and 0.5cm of qqw] (q4) {\(q_4\)} ;
			\vertex [right=0.7cm of qqw] (q4) {\(s\)} ;
			\vertex [below right=0.5cm and 0.5cm of qqw] (w2) {\(W^+\)} ;	
			
			\diagram* {
				(q1)  -- [fermion]      (qqg),
				(q2)  -- [anti fermion] (qqg),
				(qqg) -- [gluon]        (gqq),
				(gqq) -- [anti fermion] (qqa),
				(qqa) -- [anti fermion] (q3),
				(qqa) -- [boson] (aww),
				(aww) -- [boson] (w1),
				(aww) -- [boson] (wqq),
				(wqq) -- [fermion] (q5),
				(wqq) -- [anti fermion] (q6),				
				(gqq) -- [fermion] (qqw),
				(qqw) -- [fermion] (q4),
				(qqw) -- [boson] (w2),
			};
		\end{feynman}
	\end{tikzpicture}
	}
\caption{
Representative real-emission diagrams for several partonic subprocesses contributing to the EW reaction $pp \to \wpwpjjj$. 	
}
\label{fig:FD_real}
\end{figure}

%%%%%%%
Real-emission corrections to $\wpwpjjj$ production are obtained by adding an extra gluon to the subprocesses already occurring at LO, like 
\begin{align}
	q_1q_2 \to W^+W^+ q_3q_4 gg \,,\; gq_2 \to W^+W^+ q_3q_4\bar{q_1} g \,, \; 
	gg \to W^+W^+ q_3 q_4\bar q_1 \bar{q_2} \,,
\end{align}
and by subprocesses with six (anti-)quarks, such as 
\begin{align}
	\label{eq:pure-quark}
	q_1q_2 \to W^+W^+ q_3q_4 q_5\bar{q_6} \;.
\end{align}
In each case, all crossing-related combinations are taken into account. 
In Fig.~\ref{fig:FD_real} some representative diagrams retained (\ref{fig:real_gg_0res},\ref{fig:real_qq_0res}) 
or dropped (\ref{fig:real_gg_1res}, \ref{fig:real_gg_3res}, \ref{fig:real_qq_1res}, \ref{fig:real_qq_2res}) within the VBS approximation are shown. 

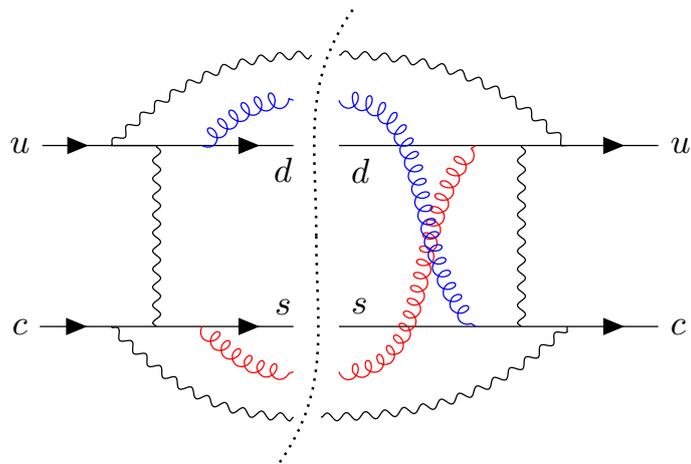
\begin{figure}%[tbp]
	\centering
	\scalebox{1.2}{
		\begin{tikzpicture}[baseline=(current bounding box)]
			\begin{feynman}
				\vertex (q1)  {\(u\)};
				\vertex [right=1cm of q1] (wu) ;
				\vertex [right=0.5cm of wu] (zu) ;
				\vertex [right=0.5cm of zu] (gu) ;
				\vertex [right=3cm of q1] (q3) ;
				\vertex [below=2cm of q1] (q2)  {\(c\)};
				\vertex [right=1cm of q2] (wd) ;
				\vertex [right=0.5cm of wd] (zd) ;
				\vertex [right=0.5cm of zd] (gd) ;
				\vertex [right=3cm of q2] (q4) ;
				\vertex [above=0.5cm of q3] (g1);
				\vertex [below=0.5cm of q4] (g2);
				\vertex [above right=0.5cm and 0.2cm of g1] (w1) ;
				\vertex [below=0.5cm of g2] (w2) ;
				\vertex [below right=1cm and 3.25cm of q1] (smid);
				\vertex [above right=2.5cm and 0.4cm of smid] (sup);
				\vertex [below left=2.5cm and 0.4cm of smid] (sdown);
				%
				% second half
				%
				\vertex [right=3.5cm of q1] (q1r) ;
				\vertex [right=3.5cm of q2] (q2r) ;
				\vertex [right=4cm of q3] (q3r)   {\(u\)};
				\vertex [right=4cm of q4] (q4r)   {\(c\)};
				\vertex [right=1.5cm of q1r] (gur) ;
				\vertex [right=0.5cm of gur] (zur) ;
				\vertex [right=0.5cm of zur] (wur) ;
				\vertex [right=1.5cm of q2r] (gdr) ;
				\vertex [right=0.5cm of gdr] (zdr) ;
				\vertex [right=0.5cm of zdr] (wdr) ;
				\vertex [right=0.5cm of g1] (g1r);
				\vertex [right=0.5cm of g2] (g2r);
				\vertex [right=0.3cm of w1] (w1r) ;
				\vertex [right=0.3cm of w2] (w2r) ;
				%
				% Only for labels
				\vertex [below right=0.0075cm and 0.00001cm of q1r] (q1rl)  {\(d\)};
				\vertex [above right=0.0075cm and 0.00001cm of q2r] (q2rl)  {\(s\)};
				%\vertex [below=0.0075cm of q1r] (q1rl)  {\(d\)};
				%\vertex [above=0.0075cm of q2r] (q2rl)  {\(s\)};
				
				\diagram* {
					(q1)   -- [fermion] (wu),
					(wu)   -- [plain]   (zu),
					(wu)   -- [plain]   (gu),
					(gu)   -- [fermion, edge label'=\(\quad\quad d\)] (q3),
					(zu)  -- [boson] (zd), 
					(q2)   -- [fermion] (wd),
					(wd)   -- [plain]   (zd),
					(zd)   -- [plain]   (gd),
					(gd)   -- [fermion, edge label=\(\quad\quad s\)] (q4),
					(wu)   -- [boson, out=45,in=180] (w1),
					(wd)   -- [boson, out=-45,in=180] (w2),				
					(gu)   -- [gluon, out=45, in=180,blue] (g1),
					(gd)   -- [gluon, out=-45, in=180,red] (g2),
					(smid) --  [ghost,out=90, in=-125] (sup),
					(sdown) -- [ghost,out=55, in=-90] (smid),
					(q1r)  -- [plain] (gur),
					(gur)  -- [plain]   (zur),
					(zur)  -- [plain]   (wur),
					(wur)  -- [fermion] (q3r),
					(q2r)  -- [plain] (gdr),
					(gdr)  -- [plain]   (zdr),
					(zdr)  -- [plain]   (wdr),
					(wdr)  -- [fermion] (q4r),
					(zur)  -- [boson] (zdr), 
					(g2r)  -- [gluon,out=0,in=-135,red]  (gur) ,
					(g1r)  -- [gluon,out=0,in=135,blue] (gdr) ,
					(w1r)  -- [boson,out=0,in=135] (wur),
					(w2r)  -- [boson,out=0,in=-135] (wdr)
%					(qgq2) -- [plain] (wd),
%					(wd)   -- [fermion] (q4),
%					(wu)   -- [boson] (wd),
%					(wd)   -- [fermion] (q4),
%					(wc)   -- [boson] (w1),
%					(wc)   -- [boson] (w2),					
				};
			\end{feynman}
		\end{tikzpicture}
		} % scalebox
	\caption{Example of a color-suppressed interference contribution to the subprocess $u c \to \wpwp d s g g$. 
	}
	\label{fig:FD_realint}
\end{figure}

We note that the consistent usage of the VBS approximation requires us to disregard colour-suppressed contributions arising from the interference of real-emission diagrams with gluons attached to different fermion lines, as depicted in Fig.~\ref{fig:FD_realint}. For the related case of VBF-induced $Hjjj$ production, this technique was described in detail in \cite{Figy:2007kv}. We follow that strategy. 

%%%%%%%%%%%%%%%%%%%%%%%%%%%%%%%%%%%%%%%%%%%
\subsection{Implementation in the \PBOXRES}\label{sec:implementation}
%%%%%%%%%%%%%%%%%%%%%%%%%%%%%%%%%%%%%%%%%%%
%
In order to perform the matching of our NLO-QCD calculation for the $\wpwpjjj$ process with PS programs we make use of the 
\PBOX{} package~\cite{Alioli:2010xd}. This program provides all process-independent elements for the matching of a fixed order calculation 
to a PS program following the \POWHEG{} method~\cite{Frixione:2007vw}, 
while process-specific input has to be provided by developers. 
This includes the flavour structures of the contributing 
partonic subprocesses, their hard-scattering matrix elements squared at LO, the virtual corrections and real-emission amplitudes squared, 
as well as spin- and colour-correlated amplitudes for the construction of infrared
subtraction terms according to the FKS scheme~\cite{Frixione:1995ms}. 

For our implementation of the VBS-induced $\wpwpjjj$ production in the framework of the resonance aware version of the \PBOX{}~\cite{Jezo:2015aia}, we employ existing tools tailored to our needs. 
The tree-level matrix-elements emerging at LO and in the real-emission contributions were generated with the \MG{} package~\cite{Alwall:2014hca} and adapted to comply with the VBS approximation.  Spin- and colour-correlated amplitudes were also constructed using \Madgraph-generated code. 
The virtual corrections were generated using \texttt{MadLoop5}~\cite{Hirschi:2011pa,Hirschi:2011rb}, supplemented by \texttt{COLLIER}~\cite{Denner:2016kdg} and \texttt{CutTools}~\cite{Ossola:2007ax} for the handling of tensor integrals and the computation of scalar integrals.

In contrast to the $\wpwpjj$ case, the VBS-induced $\wpwpjjj$ cross section exhibits singularities already at the lowest order related to the emission of soft/collinear gluons. 
For the numerical treatment of these contributions 
we employed a \textit{Born suppression factor} $F(\Phi_n)$ as introduced in~\cite{Alioli:2010qp}, using the functional form 
\begin{align}
\label{eq:bsupp}
	F(\Phi_n) = \exp\left\{ -\Lambda_1^4 \left(
	\sum_{j=1}^3     \left( \frac{1}{ p_{T,j}^2 } \right)^2
	+ \sum_{i\neq j}^3 \left( \frac{1}{ (p^2_\text{T,rel}(i,j) } \right)^2
	\right) \right \}
	\cdot \left( \frac{ h^2 }{ h^2 + \Lambda_2^2 } \right)^2		
	\; ,
\end{align}
which at each $n$-particle phase-space point $\Phi_n$ depends on the momenta of the corresponding Born configuration and the parameters $\Lambda_1$, $\Lambda_2$. 
The summation indices in \refeq{eq:bsupp} run over the three final-state partons,  $p_\text{T,rel}(i,j)$ is the relative transverse momentum of partons
$i$ and $j$ in the partonic centre-of-mass frame, and the quantity 
\begin{align}
	h=\left( \sum_{j=1}^3 p_{T,j}^2\right) \; 
\end{align}
is computed from the transverse momenta $p_{T,j}$ of the three final-state partons. 
For our numerical analyses we set the free parameters in \refeq{eq:bsupp} to $\Lambda_1 = 10$ GeV and $\Lambda_2 = 30$ GeV. 
We checked that different values of $\Lambda_{1/2}$ lead to compatible results. 
We note that an exponential suppression factor of this form was designed for the simulation of trijet production in the framework of the \PBOX{}~\cite{Kardos:2014dua}, and has also been employed in the VBF-induced $Hjjj$ process~\cite{Jager:2014vna} which features a similar singularity structure as the $\wpwpjjj$ process considered here. 

\begin{figure}%[tbp]%[h]
	\centering
	% f -> f a*
	\scalebox{1.25}{
	\begin{tikzpicture}
		\begin{feynman}
			\vertex (q1) {\(u\)};
			\vertex [right=1.5cm of q1] (qqa);
			\vertex [below right=0.5cm and 2cm of q1] (aqq);
			\vertex [below=2cm of q1] (q2) {\(c\)};
			\vertex [above right=0.5cm and 4.25cm of q1] (q3) {\(u\)};
			\vertex [above right=0.25cm and 2cm of aqq] (q4) {\(s\)};
			\vertex [right=2cm of q2]  (qgq);
			\vertex [right=1cm of qgq] (gqq);
			
			\vertex [above=1cm of qgq] (qqw);
%			\vertex [above=1cm of qqw] (aqq);
			
%			\vertex [above right=0.3cm and 0.7cm of gqq] (q5) ;
%			\vertex [below right=0.2cm and 0.7cm of gqq] (q6) ;
			\vertex [right=1cm of gqq] (q5) {\(s\)};
			\vertex [below=0.7cm of q5] (q6) {\(\bar{c}\)};
			
			\vertex [above right=0.2cm and 2cm of qqw] (w1) {\(W^+\)};
			
			\vertex [right=0.5cm of gqq] (qwq);
			\vertex [above right=0.5cm and 0.5cm of qwq] (w2) {\(W^+\)};
		
			\diagram* {
				(q1)  -- [fermion] (qqa),
				(qqa) -- [fermion] (q3),
				(aqq) -- [boson, edge label=\(\gamma^*\)] (qqa),
				(aqq) -- [fermion] (q4),
				(q2) -- [fermion] (qgq),
				(qgq) -- [gluon] (gqq),
				(gqq) -- [plain] (qwq),
				(qwq) -- [boson] (w2),
				(qwq) -- [plain] (q5),
				(q6)  -- [fermion] (gqq),
				(qgq) -- [plain] (qqw),			
				(qqw) -- [plain] (aqq),
				(qqw) -- [boson] (w1),
			};
		\end{feynman}
	\end{tikzpicture}
	}
\caption{Feynman diagram with a potentially divergent $q \to q \gamma^*$ splitting. 	}
\label{fig:FD_qqa}
\end{figure}
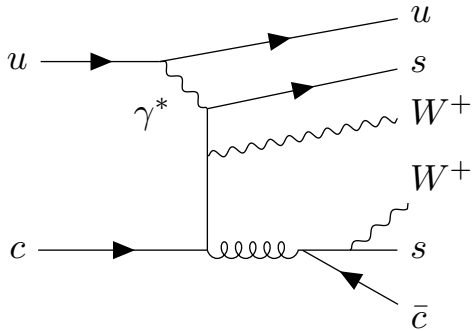

Some pure quark subprocesses of \refeq{eq:pure-quark} exhibit diagrams in which an initial-state quark couples to a virtual photon. An example is shown in Fig.~\ref{fig:FD_qqa}. Such a $q\to q\gamma^*$ splitting contains an initial-state collinear divergence that would, in principle, be cancelled by an appropriate collinear counterterm proportional to a Born-level matrix element with an initial-state photon. However, our setup assumes photon-induced contributions to be strongly suppressed and excludes initial-state photons. 
We thus regulate these divergences in the real-emission contributions by requiring that the photon virtuality be finite. Specifically, we require that $(p_i-p_f)^2 > Q^2_{\gamma,\text{min}}$, 
where $p_i$ and $p_f$ denote the four-momenta of a pair of quarks that can give rise to such a divergence. This technical cut is applied only for the all-quark real-emission subprocesses, since they are the only ones that can exhibit such divergent splittings.  
Formally, the cut introduces a dependence of the calculation upon  $Q^2_{\gamma,\text{min}}$. Nevertheless, the phase-space region considered in our numerical study (see Sec.~\ref{sec:results}) strongly disfavours configurations with very small photon virtuality. We verified that our numerical results are insensitive to the specific value of this cut in the range $Q^2_{\gamma,\text{min}}=1$ to 10~GeV$^2$. 
We remark that a method to integrate real amplitudes with divergences from soft and/or collinear photons without using a cutoff has been employed in Ref.~\cite{Gavardi:2022ixt} for diphoton production. 
There, a suppression factor was employed instead of constraining the phase-space. The method relies on the fact that the divergences are caused by external photons, which is the case in diphoton production but not in the process at hand. 
Therefore, we do not employ that method.
 
%%%%%%%%%
%
In order to validate our implementation we performed a number of comparisons and consistency checks. 
Noting that the LO contributions for $\wpwp$ production in association with three jets are identical to the real-emission corrections of the related process with two jets, we compared our new implementation to the existing one of~\cite{Jager:2011ms} for $pp\to e^+\nu_e\mu^+\nu_\mu jj$ via same-sign VBS. 
To that end we adapted the \PBOX{} implementation of~\cite{Jager:2011ms} to mimic our setup with on-shell $W^+$ bosons and found agreement between the LO contributions of our calculation and the real-emission corrections of~\cite{Jager:2011ms} for a representative setup. 

Furthermore, we performed a comparison at the level of individual Born, real-emission, and virtual amplitudes with corresponding matrix elements generated using the \recolatwo{} program~\cite{Denner:2017bdv,Actis:2016mpe}. Since the VBS approximation cannot be incorporated in \recola{}, for this comparison we concentrated on partonic subprocesses that per se only include $t$-channel exchange diagrams, such as $u c\to\wpwp d s g $.

To test the consistent implementation of real-emission amplitudes and the spin- and colour-correlated amplitudes needed for the construction of FKS subtraction terms internally within the \PBOX{} we checked that the relevant soft and collinear limits are approached correctly.

%%%%%%%%%%%%%%%%%%%%%%%%%%%%%%%%%%%%%%%%%%%
\section{Numerical analysis}\label{sec:results}
%%%%%%%%%%%%%%%%%%%%%%%%%%%%%%%%%%%%%%%%%%%
%
\subsection{Input parameters} \label{sec:input}
The results presented in this work for $\wpwpjjj$ production at the LHC have been obtained for 
proton-proton collisions at a centre-of-mass energy of $13.6$ TeV. We work in the five-flavour scheme
and use the \texttt{NNPDF40\_nnlo\_as\_01180} set 
of parton distribution functions (PDF)~\cite{NNPDF:2021njg} 
as implemented in the LHAPDF6 library~\cite{Buckley:2014ana}. 
Throughout we assume a perfect $b$-jet veto, and thus drop all contributions with $b$-quarks in the final state. 
For the Cabibbo-Kobayashi-Maskawa matrix we adopt a diagonal form.
For the EW input parameters we use the $G_\mu$ scheme where, besides the
Fermi constant $G_\mu$, the masses of the $Z$ and $W$ bosons are
fixed. For our study we choose the following input
values~\cite{ParticleDataGroup:2022pth}:
\begin{equation}
    m_Z = 91.1876~\mr{GeV}\,,\quad 
    m_W = 80.377~\mr{GeV}\,,\quad 
    G_\mu = 1.1663788\times 10^{-5}~\mr{GeV}^{-2}\,.
\end{equation}
The EW coupling $\alpha$ and the weak mixing angle 
$\theta_W$  are derived from these via tree-level relations. The widths of the $Z$ and $W$ bosons
are set to: 
\begin{equation}
    \Gamma_Z = 2.4955~\mr{GeV}\,,\quad 
    \Gamma_W = 2.085~\mr{GeV}\,, 
\end{equation}
and the top-quark mass and width to: 
\begin{equation}
    m_t = 172.5~\mr{GeV}\,, \quad
    \Gamma_t = 1.42~\mr{GeV}\,. 
\end{equation}
For the mass and width of the Higgs boson we employ:
\begin{equation}
    m_H = 125.25~\mr{GeV}\,, \quad
    \Gamma_H = 3.2\times 10^{-3}~\mr{GeV}\,.
\end{equation}
The minimal photon virtuality for the technical cut discussed in Sec.~\ref{sec:implementation} is set to $Q^2_{\gamma,\text{min}} = 4$ GeV$^2$. 

We define the renormalization and factorization scales as multiples of
a central scale $\mu_0$ according to $\mur=\xir\,\mu_0$ and
$\muf=\xif\,\mu_0$ with scale factors $\xi_R$ and $\xi_F$, and consider 
the dynamical central scale 
\begin{equation}\label{eq:dynamical-scale}
	\mu_0 = \dfrac{1}{2}\left(E_{T,W_1}+E_{T,W_2}+ \sum_f^{\mr{n_{part}}} p_{T,f}\right)\,, 
\end{equation}
where
\begin{equation}
	E_{T,W_i} = \sqrt{m_W^2+p_{T,W_i}^2} \;
\end{equation}
is the transverse energy of the on-shell $W$ boson $i$ $(i=1,2)$ with transverse momentum $p_{T,W_i}$, and
the sum in Eq.~(\ref{eq:dynamical-scale}) includes the transverse momenta $p_{T,f}$ of all
$\mr{n_{part}}$ final-state partons of a considered Born-type or
real-emission configuration.
For the parton-shower results in this paper, we performed a seven-point renormalization- and factorization-scale variation using the reweighting function of the \PBOX{} to estimate the scale uncertainty. 
To this end, we independently set $\xif$ and $\xir$ to the values $\frac{1}{2}$, $1$ and $2$ while excluding the combinations $(\xif,\xir) = (\frac{1}{2},2)$ and $(\xif,\xir) = (2,\frac{1}{2})$. 

We employ the anti-$k_T$ jet algorithm~\cite{Cacciari:2008gp} with distance parameter $R=0.4$, as implemented 
in the \texttt{FastJet} package~\cite{Cacciari:2011ma}. 

Our phenomenological analysis imposes cuts to define the phase-space region in which the VBS 
approximation is trustworthy. Three jets are required with transverse momenta and rapidities such that
\begin{align}\label{eq:cuts1}
    p_{T,j_k} \geq 30~\mr{GeV}\; , \quad |y_{j_k}| \leq 4.5 \, ,
\end{align}
with $k=1,2,3$.
We denote the hardest two of these jets as \textit{tagging jets} $j_1$ and $j_2$, 
and require that they are well separated and have a large invariant mass, 
\begin{align}\label{eq:cuts2}
	|\Delta y_{j_1,j_2}| \geq 2.5 \; , \quad M_{j_1,j_2} \geq 500~\mr{GeV} \; .
\end{align}
No cuts are imposed on the final-state gauge-boson system. 

%%%%%%%%%%%%%%%%%%%%%%%%%%%%%%%%%%%%%%%%%%%%%%%%%%%%%%%%%%%%%%%%%
\subsection{Results at fixed order}\label{sec:results_fo}
For the $\wpwpjjj$ cross sections at LO and NLO QCD integrated over the fiducial phase space within the cuts of Eqs.~(\ref{eq:cuts1})--(\ref{eq:cuts2}) we obtain $\sigma_\mr{LO} = 3.214(2)\times10^{-2}~\mr{fb}$ and
$\sigma_\mr{NLO} = 3.65(2)\times10^{-2}~\mr{fb}$, respectively, 
where the Monte Carlo integration uncertainty on the last digit is given in parentheses. 
This corresponds to a relative QCD correction on the LO result of $+13.7(9)\%$. 
We note that branching ratios for the decay of the $\wpwp$ gauge boson system are not included in these numbers.

\begin{figure}[t]%[htb]
	\centering
	\includegraphics[width=0.48\textwidth]{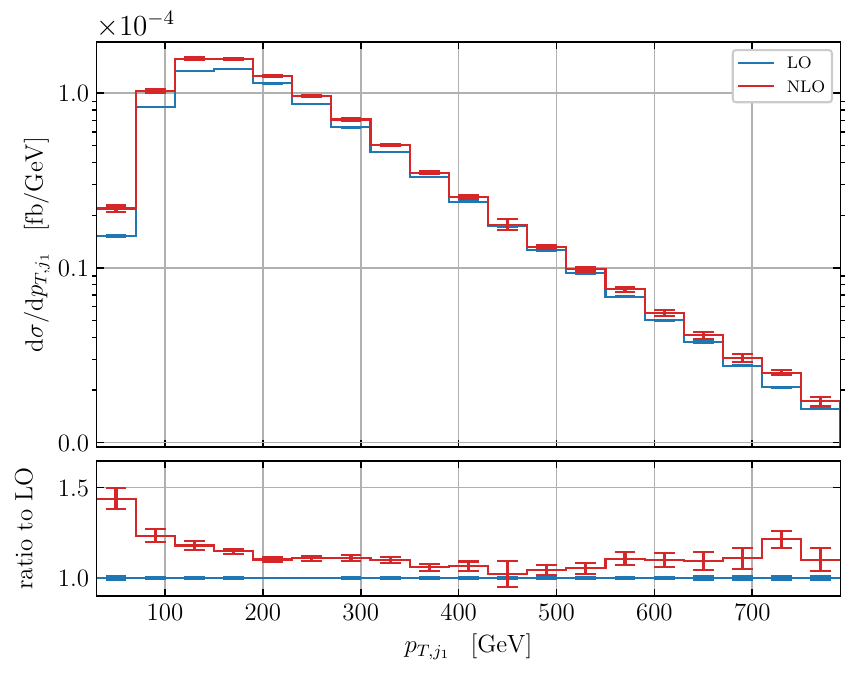}
	\includegraphics[width=0.49\textwidth]{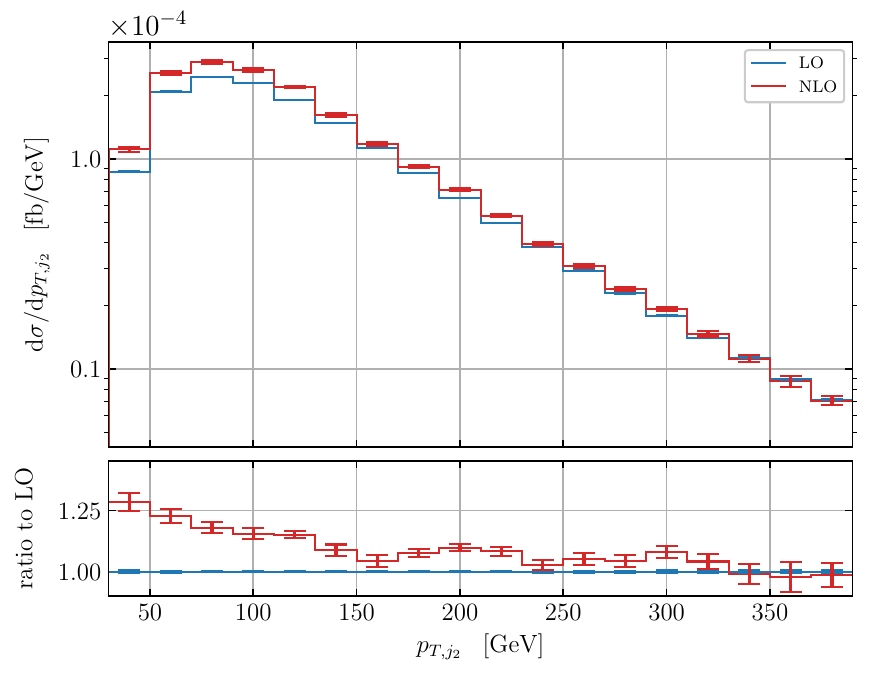}
	\includegraphics[width=0.48\textwidth]{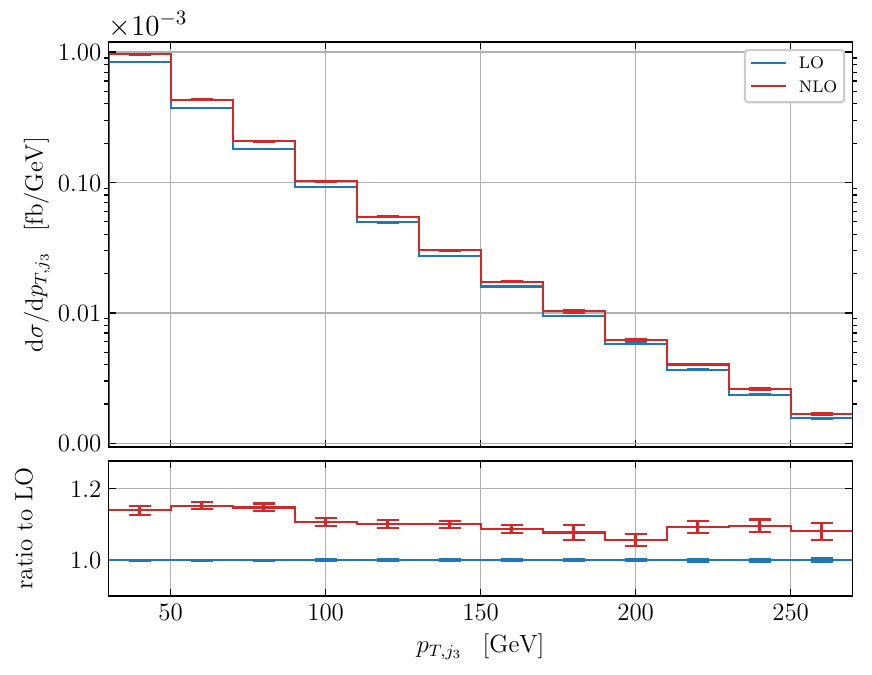}
	\includegraphics[width=0.48\textwidth]{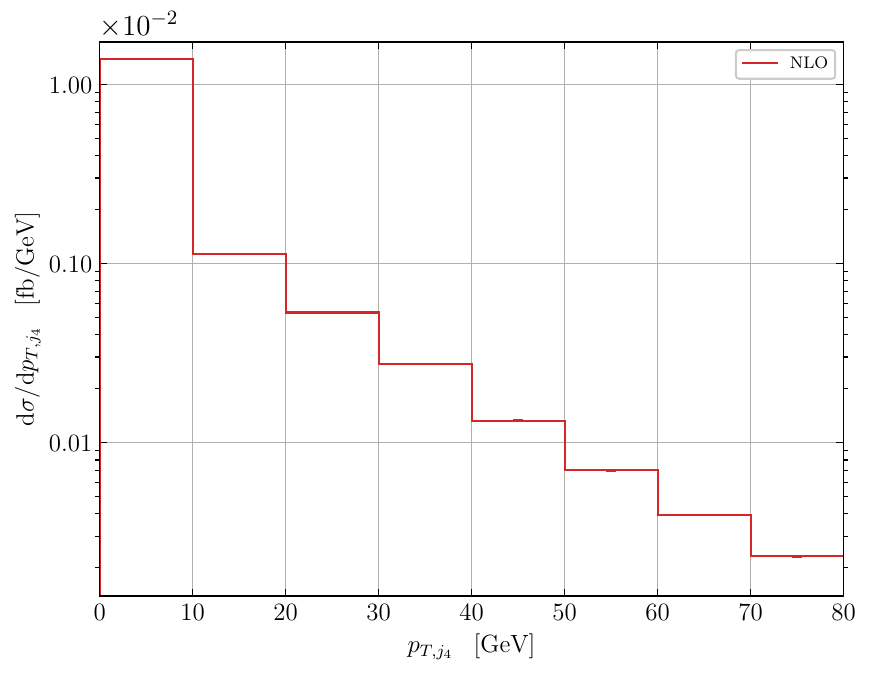}
	\caption{Transverse-momentum distributions of the four jets encountered in $pp\to \wpwpjjj$ at the LHC within the cuts of Eqs.~\eqref{eq:cuts1}-\eqref{eq:cuts2} at LO (blue) and NLO (red) in QCD (upper panels), as well as their ratios (lower panels). The error bars correspond to the Monte Carlo integration uncertainty. 
	}
	\label{fig:LO_vs_NLO-pt}
\end{figure}

\begin{figure}[t]%[htb]
	\centering
	\includegraphics[width=0.48\textwidth]{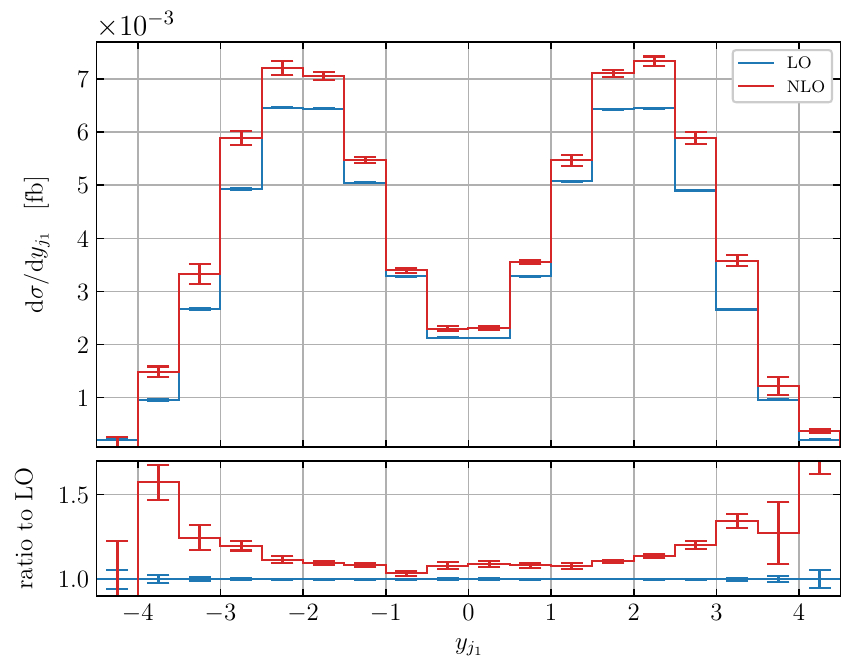}
	\includegraphics[width=0.48\textwidth]{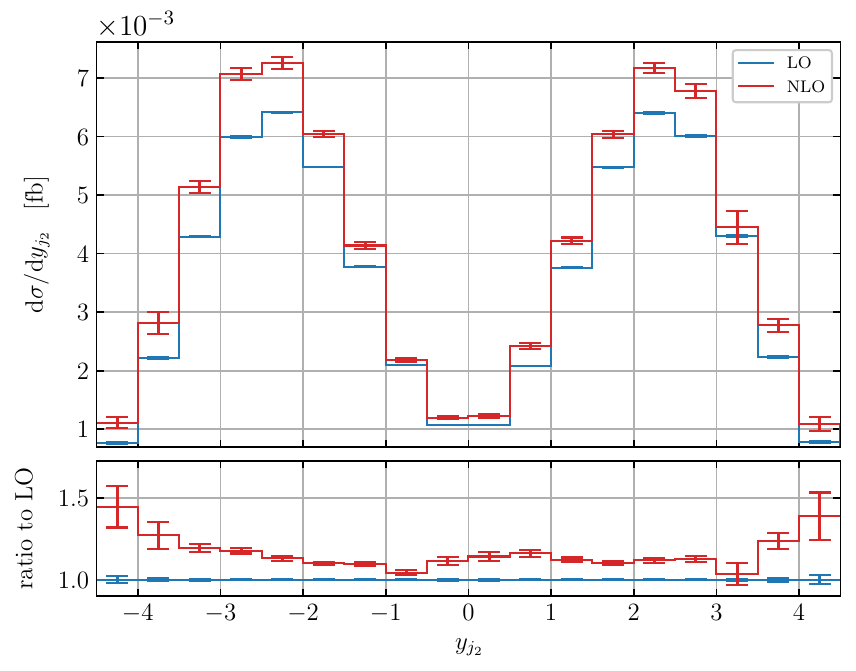}
	\includegraphics[width=0.48\textwidth]{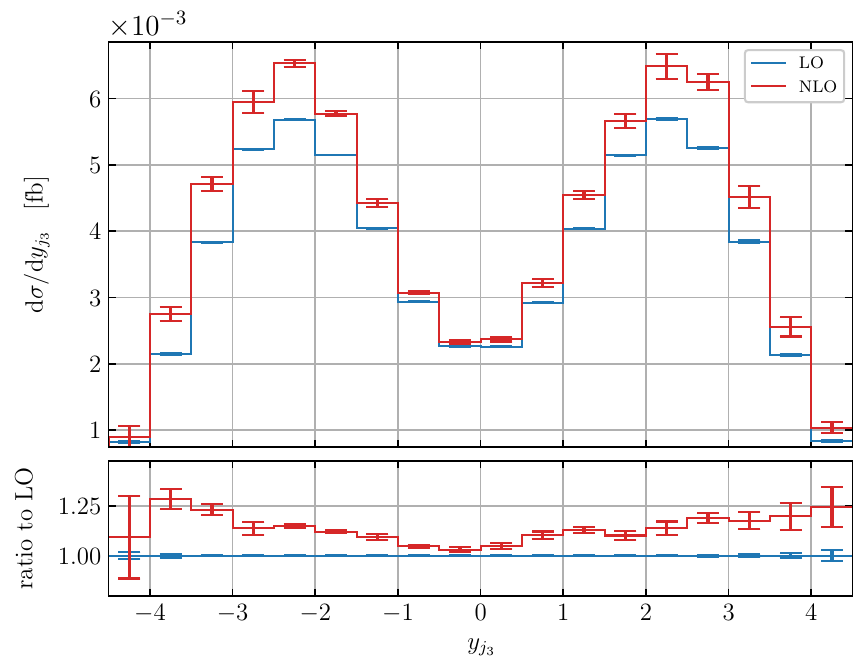}
	\includegraphics[width=0.48\textwidth]{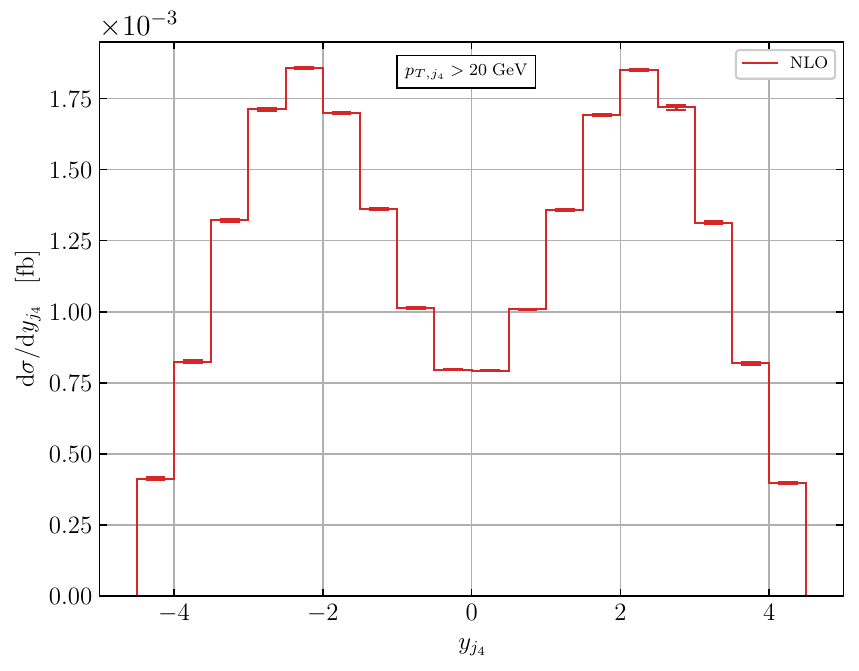}
	\caption{Similar to Fig.~\ref{fig:LO_vs_NLO-pt}, but for the rapidity distributions of the jets. For the rapidity distribution of the fourth jet the additional cut of \refeq{eq:cut-j4} is imposed. }
	\label{fig:LO_vs_NLO-y}
\end{figure}

\begin{figure}[t]%[htb]
	\centering
	\includegraphics[width=0.48\textwidth]{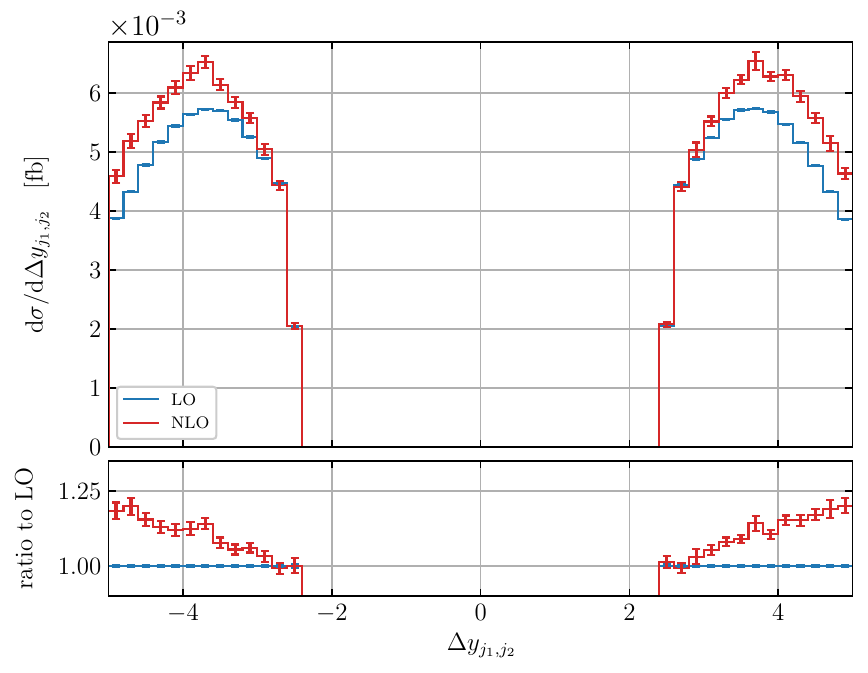}
	\includegraphics[width=0.48\textwidth]{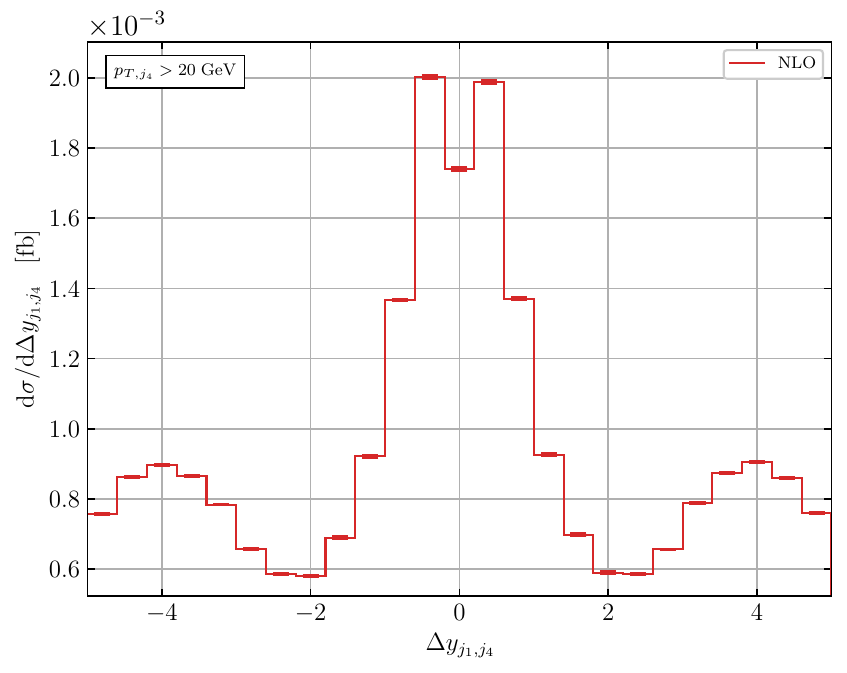}
	\includegraphics[width=0.48\textwidth]{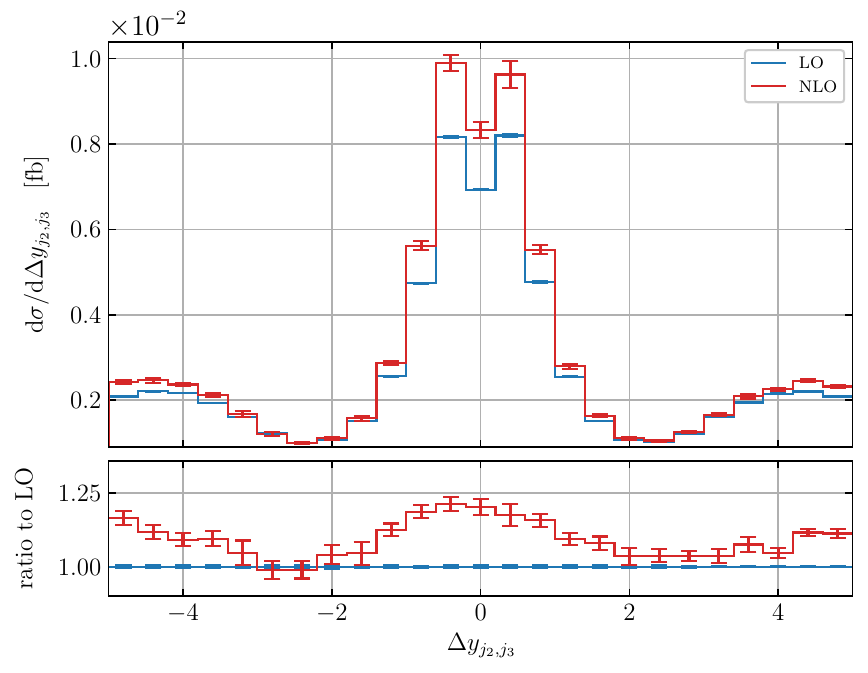}
	\includegraphics[width=0.48\textwidth]{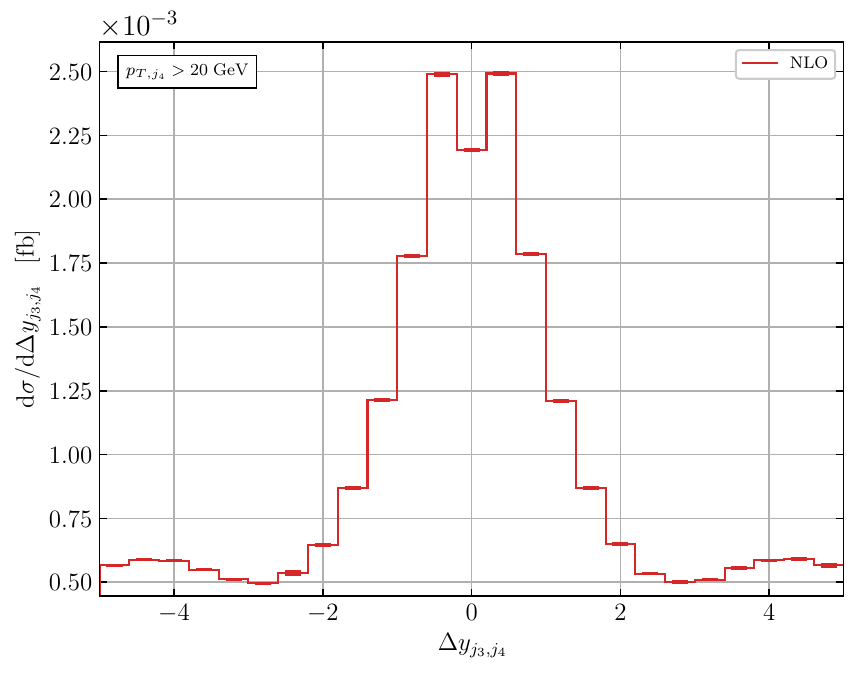}
	\caption{
	Similar to Fig.~\ref{fig:LO_vs_NLO-pt}, but for  the rapidity separations of various jet pairs. For distributions involving the fourth jet the additional cut of \refeq{eq:cut-j4} is imposed. 
	}
	\label{fig:LO_vs_NLO-dyjj}
\end{figure}

In the following, we investigate the effects of the NLO corrections for several jet observables. Figures~\ref{fig:LO_vs_NLO-pt} to \ref{fig:LO_vs_NLO-dyjj} show distributions at LO and NLO in QCD, together with their respective ratios. 
Some distributions of the fourth jet are computed assuming a lower bound on its transverse momentum,  
\begin{align}
\label{eq:cut-j4}
p_{T,j_4}\geq 20~\gev\,.
\end{align}
When present, this additional cut is specified in the respective figures.

The transverse momentum distributions of the four jets that can occur in $pp\to\wpwpjjj$ at NLO-QCD are shown in Fig.~\ref{fig:LO_vs_NLO-pt}.
The transverse momenta of the tagging jets $j_1$ and $j_2$ display maxima at 110--190 GeV and 70--90 GeV, respectively. This behaviour is typical for VBS processes (see, e.g.~\cite{Rainwater:1999sd}). The transverse-momentum distributions of the subleading jets strongly increase towards small values where soft/collinear emissions dominate. We note that at NLO in QCD a fourth jet can only stem from real-emission contributions, and is thus effectively only described with leading-order accuracy. In contrast, the third jet is accounted for with full NLO-QCD accuracy in our $\wpwpjjj$ calculation. We are thus able to describe the behaviour of subleading jets with higher precision than a calculation for the $\wpwpjj$ production process, where a third jet occurs only in real-emission corrections.  We note that a quantitative understanding of the subleading jets is of paramount importance for the design of signal-selections strategies that rely on central-jet-veto (CJV) techniques~\cite{Barger:1994zq,Rainwater:1996ud,Kauer:2000hi}. CJV analyses make use of the distinct feature of VBS processes that exhibit little jet activity in the central region because of the colour singlet nature of the $t$-channel weak-boson exchange. Consequently, when a CJV is applied events with a jet in the rapidity region between the two tagging jets are discarded. This results in a significant improvement of signal-to-background ratios, as the dominant QCD background processes  typically feature a pronounced jet activity in this very region.

Figure~\ref{fig:LO_vs_NLO-y} displays the rapidity distribution of each jet.
For the three hardest jets, the NLO corrections increase towards larger absolute values of their rapidity. Both LO and NLO rapidity distributions of all jets 
assume their maximal values between $\pm 2$ and $\pm 2.5$. This is expected for VBS processes, in which most of the included topologies correspond to forward, 
back-to-back tagging jets which radiate the third and fourth jets. This picture is compatible with the rapidity-difference distributions of Fig.~\ref{fig:LO_vs_NLO-dyjj}, 
which illustrate that non-tagging jets tend to have rapidities that are similar to each other and to those of the tagging jets, while the tagging jets exhibit a large rapidity separation. This behaviour has been observed before for the third jet in VBS processes (see, for example, Refs.~\cite{Jager:2014vna,Jager:2018cyo}). 
In detail, we find that for the $\Delta y_{j_1,j_2}$ distribution, which is restricted to absolute values larger than 2.5 because of the rapidity-separation cut of \refeq{eq:cuts1}, the LO and NLO distributions assume their maxima at $\pm3.8$. These large separations are expected for VBS processes. The NLO corrections are slightly increasing towards larger rapidity separations. 
In contrast, the rapidity separations of a non-tagging jet to one of the tagging jets and of the two non-tagging jets tend to be small with an interesting peak structure at about $\Delta y_{j_i,j_k}\approx\pm 0.4$. 
Yet smaller rapidity separations are slightly suppressed by the requirement of separately identifiable jets for our numerical simulations with an $R$ parameter of 0.4.

%%%%%%%%%%%%%%%%%%%%%%%%%%%%%%%%%%%%%%%%%%%%%%%%%%%%%%%%%%%

\subsection{Parton-shower results}
\label{sec:results_nlops}
%%%%%%%%%%
To assess PS effects, we match our calculation to \PYTHIA~\texttt{8.240}~\cite{Sjostrand:2014zea} using the \texttt{Monash 2013} tune~\cite{Skands:2014pea}. We switch off QED showering for all results. Other \PYTHIA{} settings are varied and accordingly specified below.

We begin by comparing distributions at fixed next-to-leading order (termed NLO) and NLO matched to PS (referred to as NLO+PS in the following). For this comparison, we employ the dipole recoil of the \PYTHIA{} shower. 
We deactivate multi-parton interactions (MPI) and hadronisation effects. We denote this choice of settings as the \PYDS{} setup. 
For the corresponding \NLOPS{} cross section within the selection cuts of Eqs.~\eqref{eq:cuts1}-\eqref{eq:cuts2} we obtain a value of 
$\sigma_{\PYDS}=3.45(1)\times10^{-2}~\mr{fb}$, 
which represents an 
$8.2\%$ 
reduction with respect to the fixed-order result quoted above.  
This is an expected effect, since fewer events pass the selection cuts at \NLOPS{} level due to the energy loss of the jets from the additional radiation generated by the PS. 

\begin{figure}[t]%[htb]
	\centering
		\includegraphics[width=0.48\textwidth]{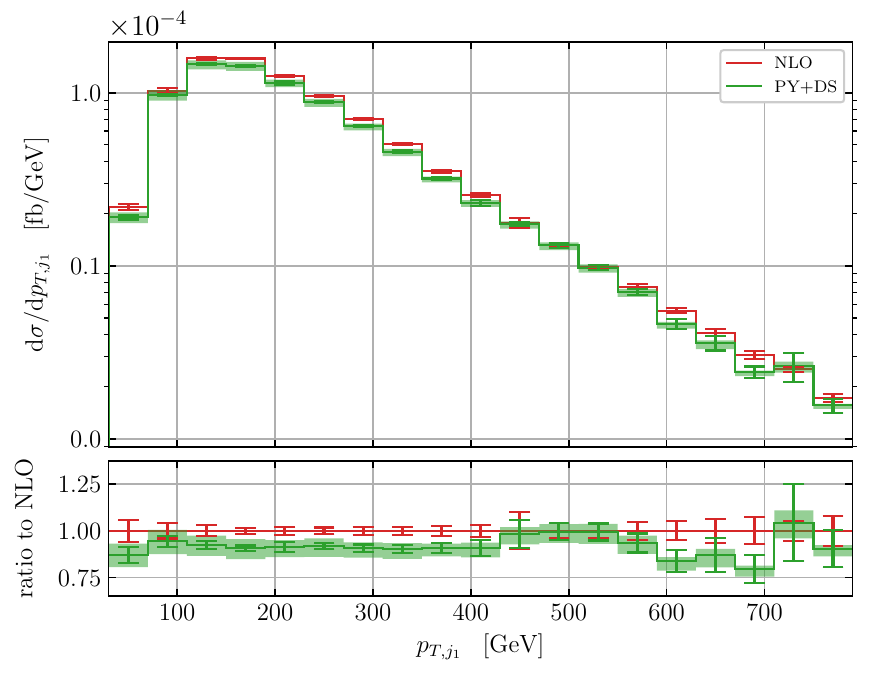}
		\includegraphics[width=0.48\textwidth]{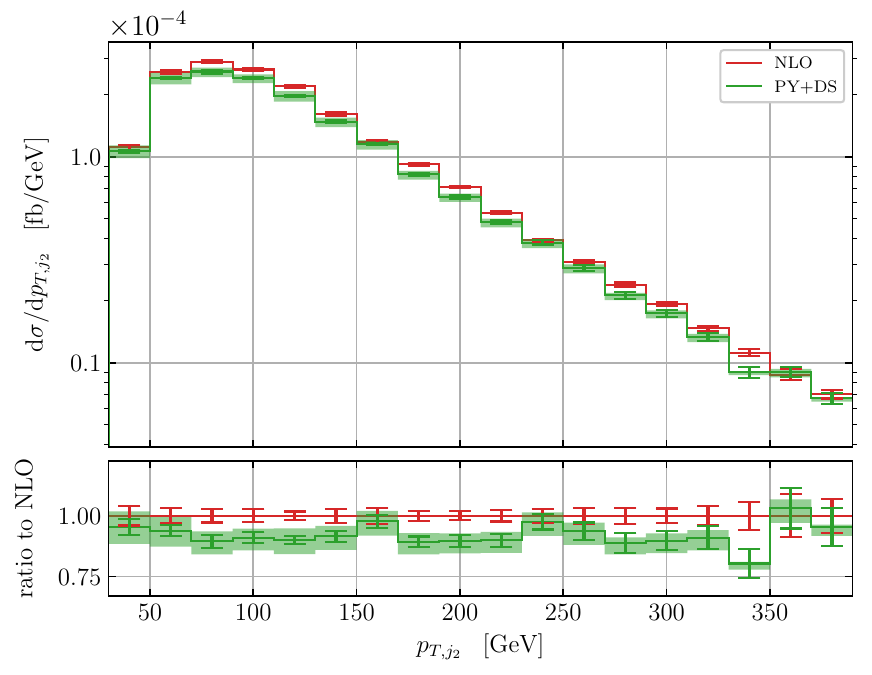}
		\includegraphics[width=0.48\textwidth]{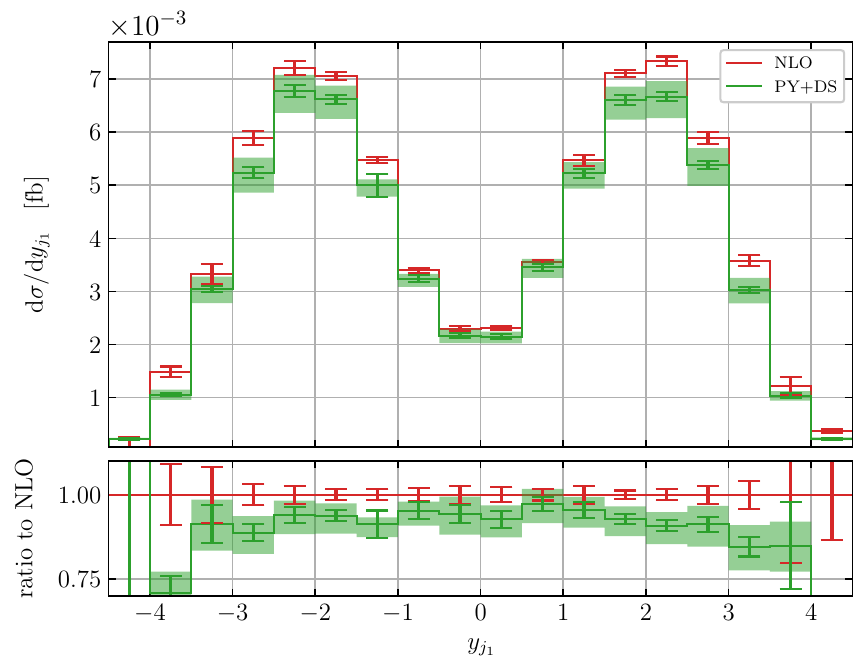}
		\includegraphics[width=0.48\textwidth]{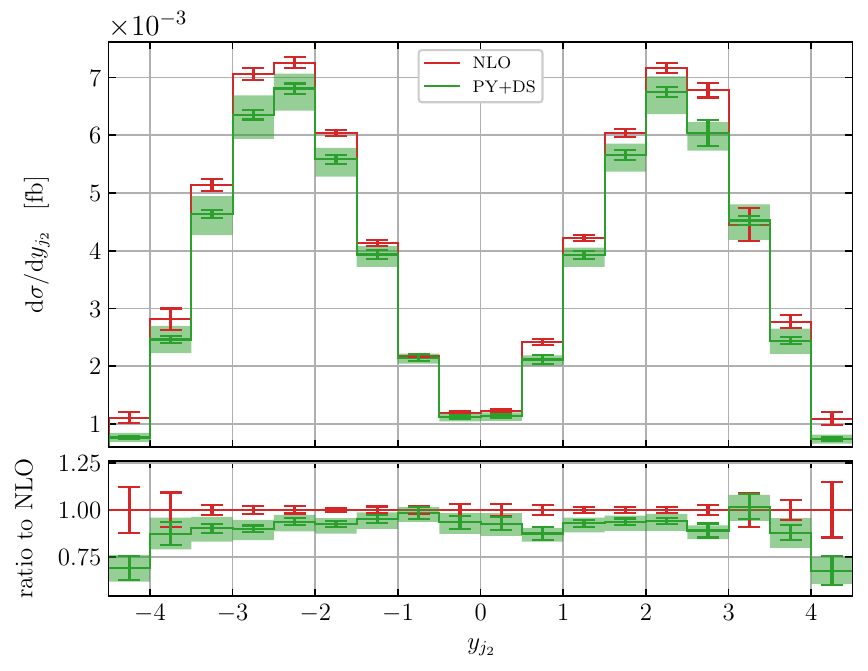}
	\caption{ Transverse-momentum and rapidity distributions of the two tagging jets encountered in $pp\to \wpwpjjj$ at the LHC within the cuts of Eqs.~\eqref{eq:cuts1}-\eqref{eq:cuts2} at  NLO (red) and \NLOPS{} (green) using the dipole shower in \PYTHIAE{} (upper panels), as well as their ratios (lower panels). The error bars indicate statistical uncertainties, while the bands indicate scale uncertainties obtained by a 7-point variation of $\muf$ and $\mur$.
	}
	\label{fig:NLO_vs_PY8-tagjets}
\end{figure}

\begin{figure}[t]%[htb]
	\centering
	\includegraphics[width=0.48\textwidth]{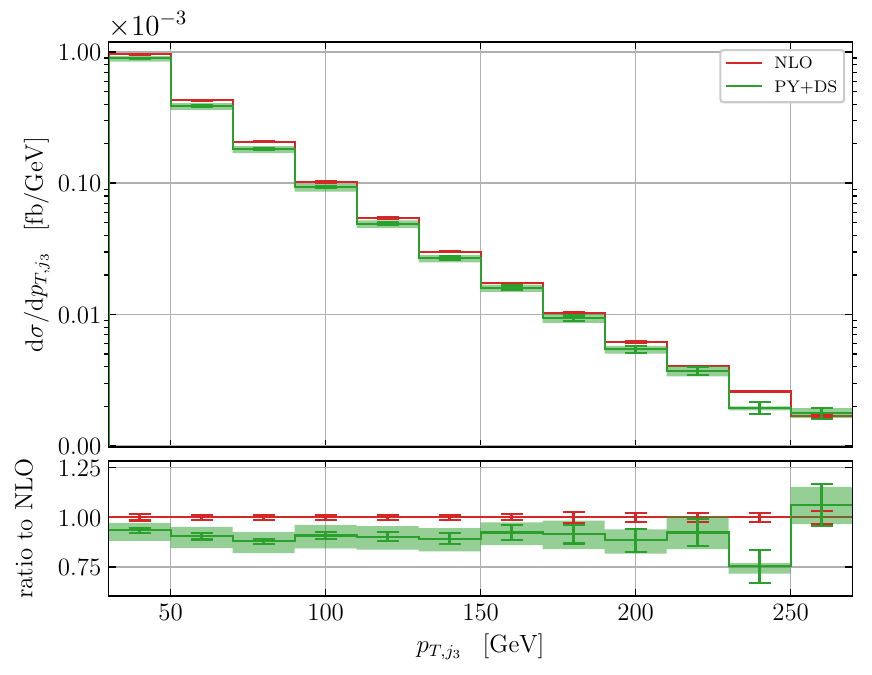}
	\includegraphics[width=0.48\textwidth]{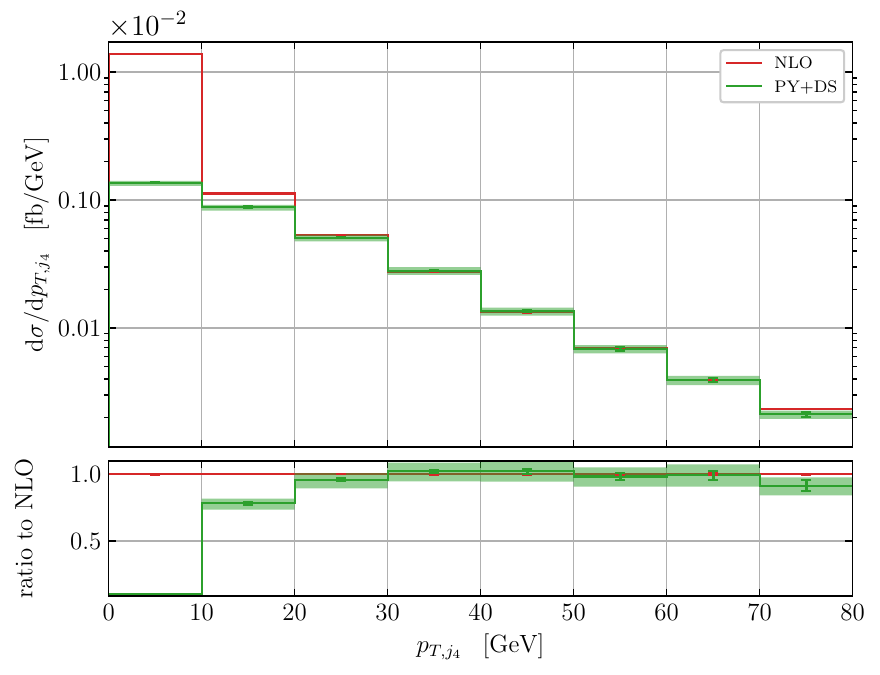}
	\includegraphics[width=0.48\textwidth]{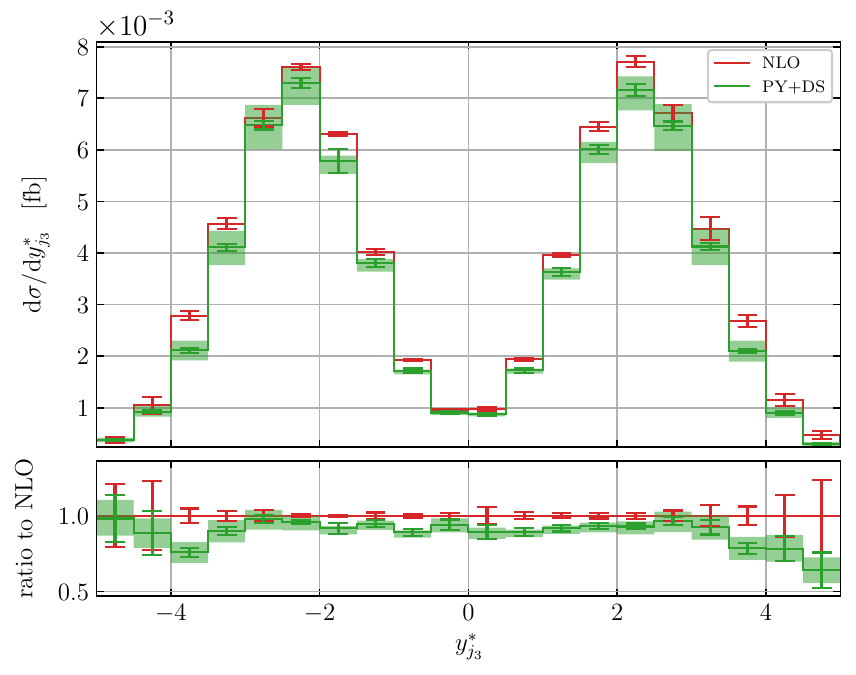}
	\includegraphics[width=0.48\textwidth]{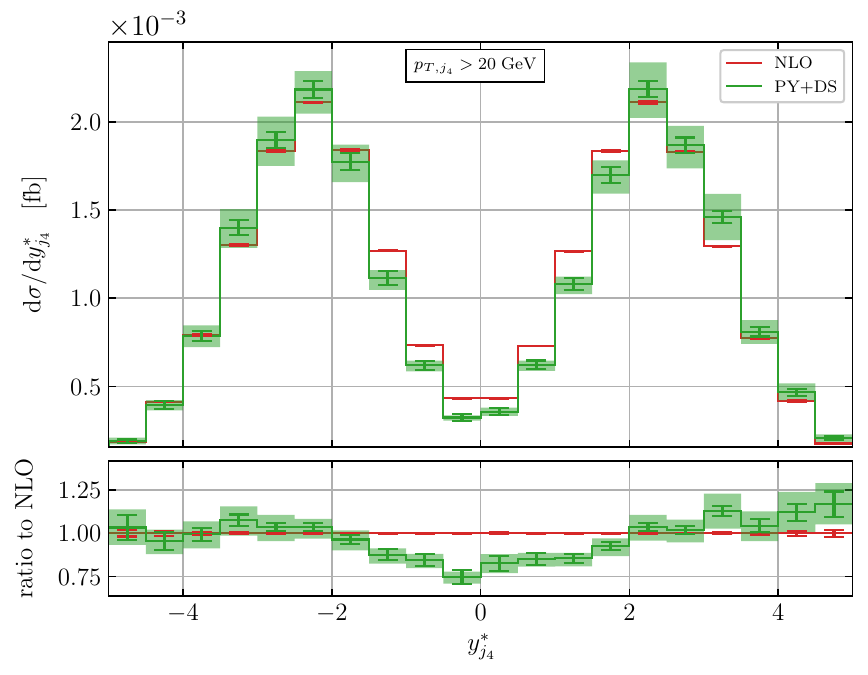}
	\caption{Similar to Fig.~\ref{fig:NLO_vs_PY8-tagjets}, but for transverse-momentum and relative-rapidity distributions of the non-tagging jets. For the  $y^*_{j_4}$ distributions the additional cut of \refeq{eq:cut-j4} is imposed.
	}
	\label{fig:NLO_vs_PY8-nontagjets}
\end{figure}

Let us now turn to a discussion of PS effects on NLO differential distributions. 
Figure~\ref{fig:NLO_vs_PY8-tagjets} shows the transverse-momentum and rapidity distributions of the tagging jets, 
revealing that the PS has little impact on these distributions in the considered setup. We will discuss below that different PS settings can change that behaviour. 
More significant PS effects occur for distributions of the non-tagging jets, 
illustrated by Fig.~\ref{fig:NLO_vs_PY8-nontagjets}. We note that, as specified in \refeq{eq:cuts1}, we have applied a transverse-momentum cut on the third jet which avoids a population of phase-space regions with low values of $p_{T,j_3}$. In the allowed regime,  the shape of its transverse-momentum  distribution is not significantly affected. On the other hand, for the fourth jet we observe the expected Sudakov damping which results in a redistribution of events of low transverse momenta to larger values.
In order to assess the separation of a non-tagging jet $j_i$ ($i=3,4$) from the tagging jets, we consider its rapidity relative to the centre of the tagging-jet system, 
\begin{align}
	y_{j_i}^* = y_{j_i} - \frac{y_{j_1} + y_{j_2}}{2} \; . 
\end{align} 
The relative rapidity of the third jet in the \PYDS{} setup is comparable to the NLO result within the estimated scale uncertainties.  However, we observe a slight tendency of the PS to  shift the fourth jet towards higher values of $y_{j_4}^*$, that is away from the central rapidity region. 

%%%%%%%%%%%%%%
%
Let us now compare different settings of the PS program: 
Instead of the dipole shower used for the \PYDS{} results, for the \PYGS{} setup we use the global recoil shower of \PYTHIAE{}. 
In the \PYMPI{} scheme the dipole recoil is used, and additionally MPI and hadronisation effects are activated. 

\begin{figure}[t]%[htb]
	\centering
	\includegraphics[width=0.48\textwidth]{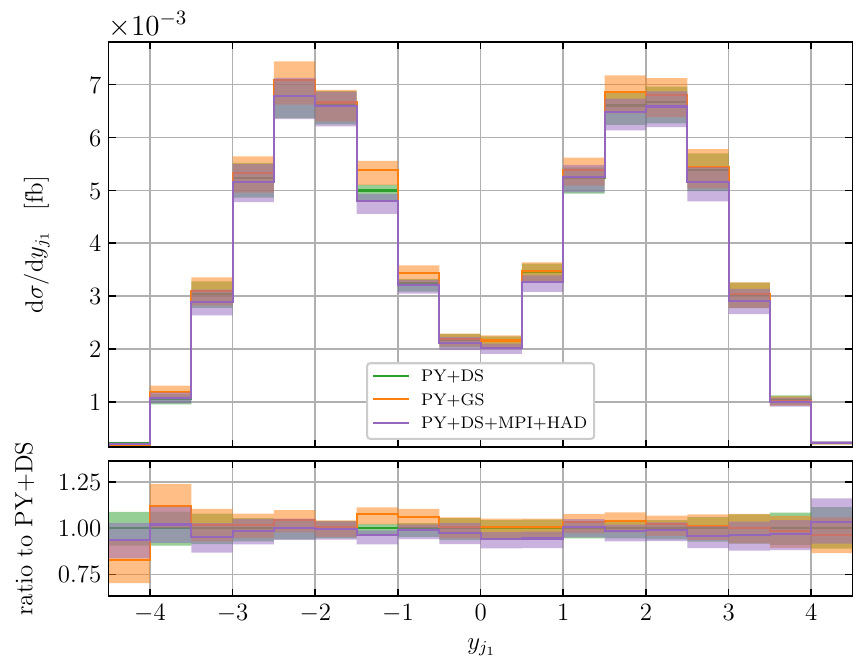}
	\includegraphics[width=0.48\textwidth]{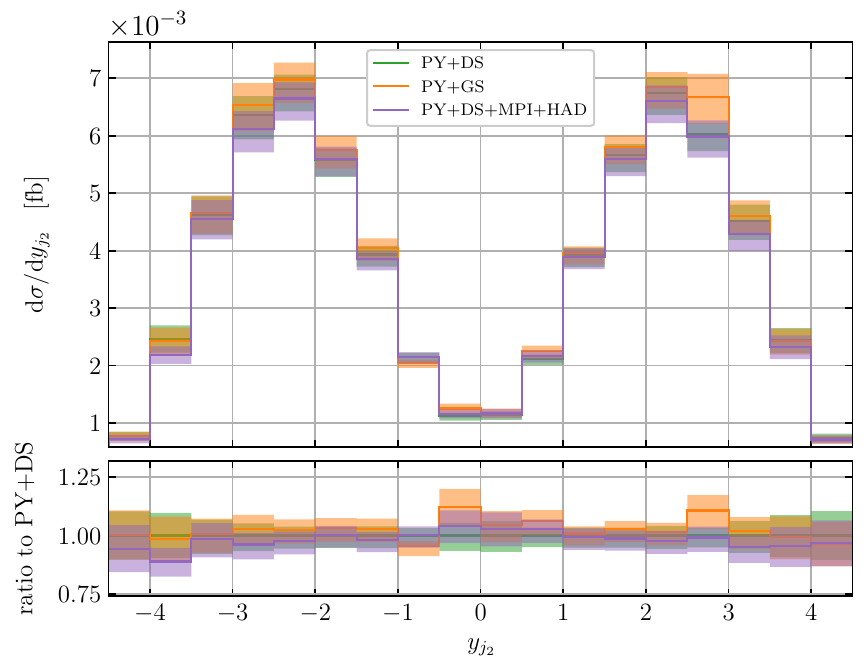}
	\includegraphics[width=0.48\textwidth]{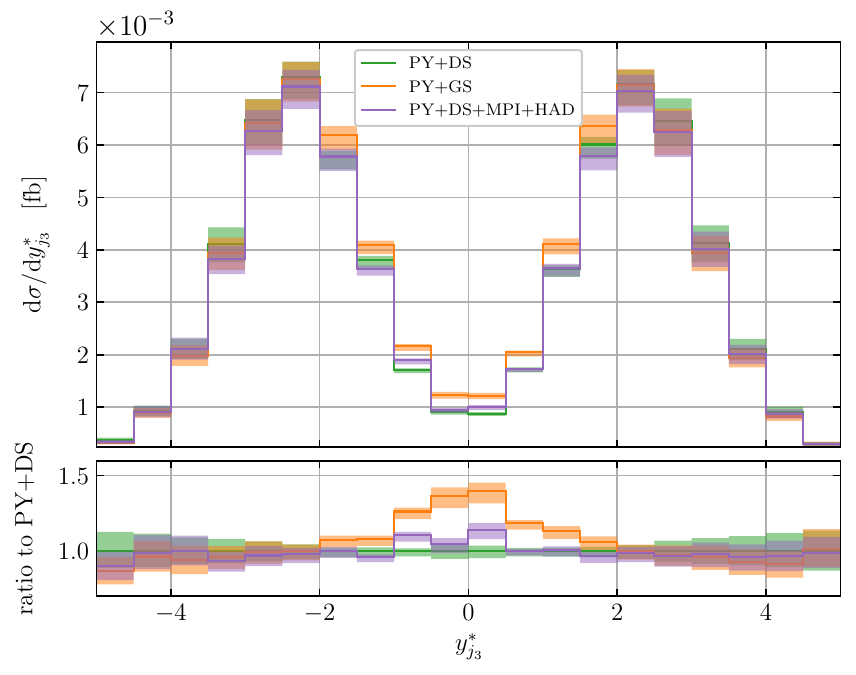}
	\includegraphics[width=0.48\textwidth]{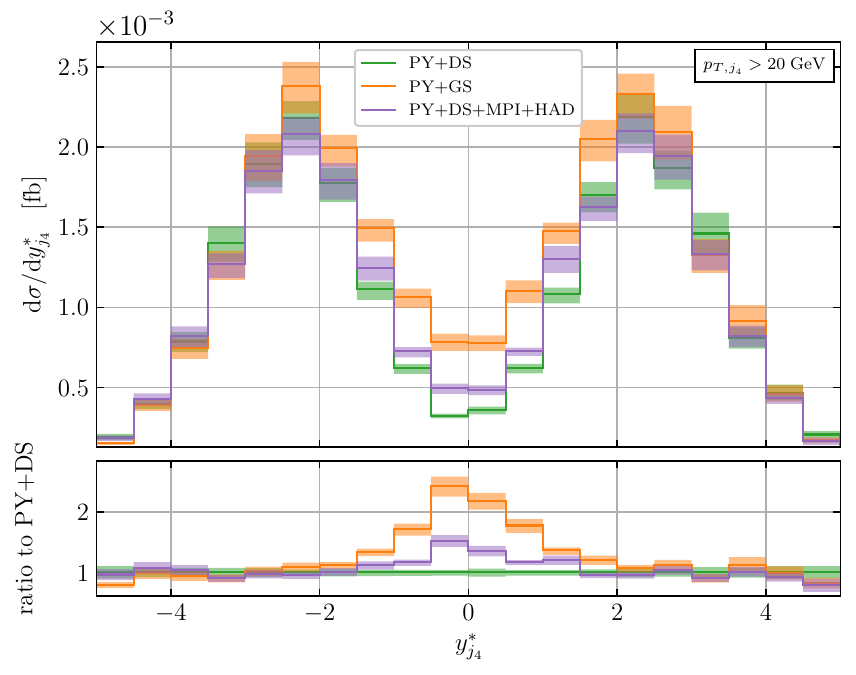}
	\caption{Rapidity distributions of the tagging jets (upper plots) and rapidity distributions of the non-tagging jets relative to the tagging jets (lower plots) encountered in $pp\to \wpwpjjj$ at the LHC within the cuts of Eqs.~\eqref{eq:cuts1}-\eqref{eq:cuts2} at  \NLOPS{} accuracy for the \PYDS{} (green), \PYGS{} (orange), and 
	\PYMPI{} (purple) setups, 
	and their ratios to the \PYDS{} results. For the  $y^*_{j_4}$ distributions the additional cut of \refeq{eq:cut-j4} is imposed. 
	The bands indicate scale uncertainties obtained by a 7-point variation of $\muf$ and $\mur$.  
	}
	\label{fig:PY8-comp}
\end{figure}

Figure~\ref{fig:PY8-comp} displays predictions for the rapidity distributions of all jets.  
While for the tagging jet distributions differences due to PS settings are basically contained within the scale uncertainties estimated by the 7-point variation of $\muf$ and $\mur$ introduced in Sec.~\ref{sec:input}, 
they are more pronounced in distributions of the non-tagging jets, 
and especially strong for the fourth jet. 
We observe that the regions of  low $\ystart$ and $\ystarf$ are more populated when the global recoil scheme is used instead of a dipole recoil scheme. Activating MPI and hadronisation effects results in a slight increase of events with low values of $\ystar$, but still causes the non-tagging jets to better reproduce the pure fixed-order results than the global recoil scheme. 
This observation is in agreement with expectations from the literature~\cite{Hoche:2021mkv}, and strengthens the case for recommending the use of a dipole recoil scheme for the simulation of VBS processes with a PS generator.  
%{\red 
Indeed, noting that distortions on observables like the rapidity of the third jet -- that is described with NLO accuracy in our calculation -- caused by the use of an inappropriate recoil scheme in the parton shower simulation are larger than the NLO corrections themselves, we strongly advocate the user to abstain from the simulation of VBS processes by means of a global recoil scheme. 
%}

%%%%%%%%%%

\subsection{Effects of a Central-Jet Veto}
We now investigate more in detail the effects of a CJV.  
Following Ref.~\cite{Figy:2007kv}, we veto an event, if any of the subleading jets $j_s$ with transverse momentum exceeding the veto value of $\ptveto$ is located in the rapidity interval between the two tagging jets, fulfilling simultaneously the conditions
\begin{align}\label{eq:cjv}
	\min\left(y_{j_1},y_{j_2}\right)  < y_{j_{s}} < \max\left(y_{j_1},y_{j_2}\right) \quad\text{and}\quad p_{T,j_{s}} > \ptveto \,.
\end{align}
Unless explicitly stated otherwise, we set the value of $\ptveto$  to 
\begin{align}\label{eq:ptveto}
\ptveto  = 20~\gev\,.
\end{align}
This veto is employed in addition to the cuts of Eqs.~(\ref{eq:cuts1})--(\ref{eq:cuts2}).

\begin{figure}[t]%[htb]
	\centering
	\includegraphics[width=0.31\textwidth]{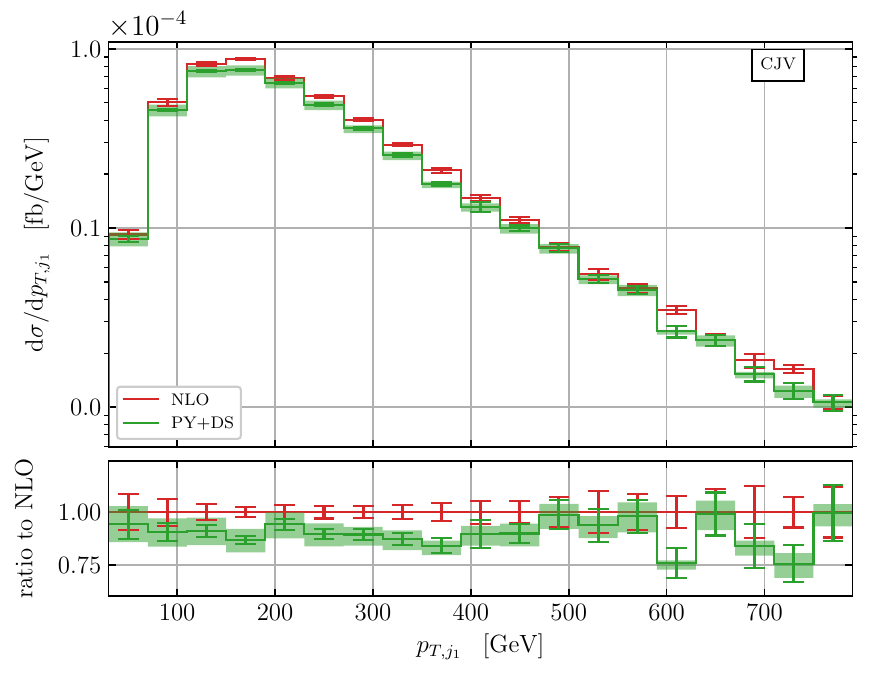}
	\includegraphics[width=0.31\textwidth]{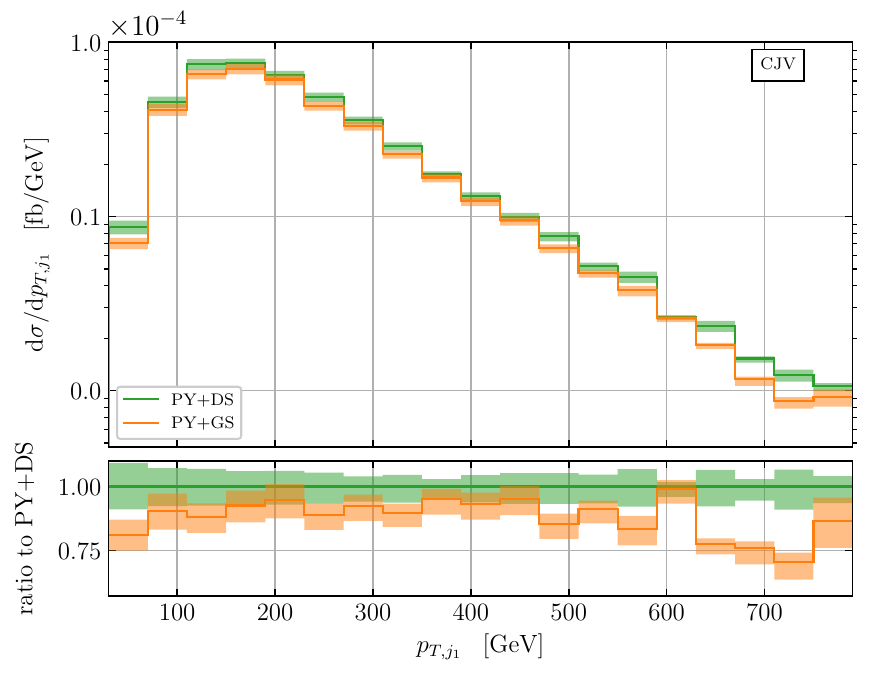}
	\includegraphics[width=0.31\textwidth]{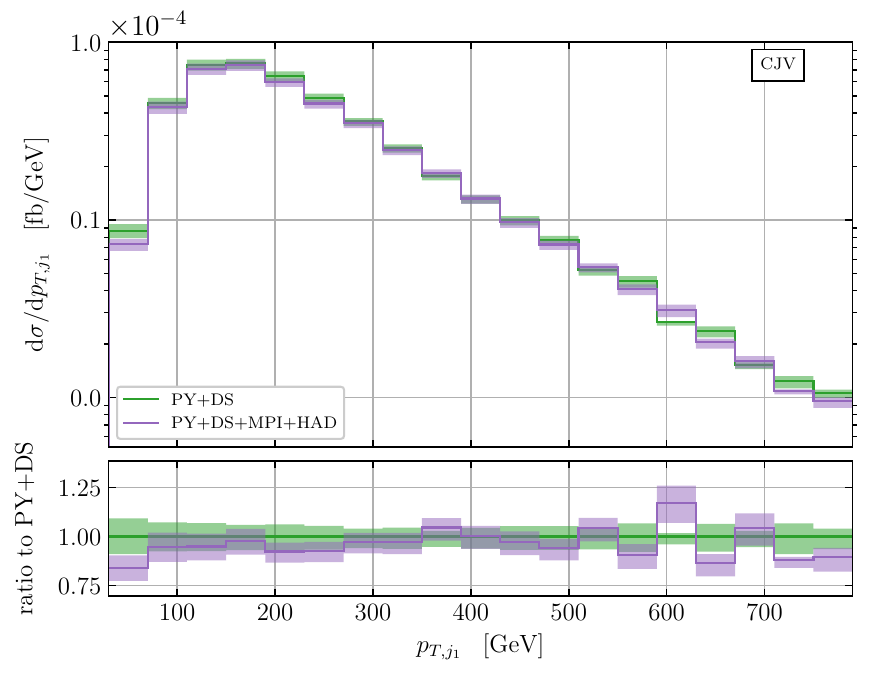}	
	\caption{Transverse-momentum distributions of the hardest tagging jet encountered in $pp\to \wpwpjjj$ at the LHC within the cuts of Eqs.~\eqref{eq:cuts1}--\eqref{eq:cuts2} and Eqs.~\eqref{eq:cjv}--\eqref{eq:ptveto} at NLO and \NLOPS{} accuracy using the \PYDS{} setup (left), for the \PYDS{} and \PYGS{} setups (center), and for the \PYDS{} and \PYMPI{} setups (right), together with their ratios.  
	In each case the bands indicate scale uncertainties obtained by a 7-point variation of $\muf$ and $\mur$.  
	}
	\label{fig:cjv-pt1}
\end{figure}

First we consider observables related to the tagging jets, such as the transverse-momentum distribution of the hardest tagging jet that has already been discussed above for the case without a CJV at fixed order (c.f.\ Fig.~\ref{fig:LO_vs_NLO-pt}) and at \NLOPS{} accuracy (c.f.\ Fig.~\ref{fig:NLO_vs_PY8-tagjets}). 
In Fig.~\ref{fig:cjv-pt1} 
we show predictions for this distribution within the cuts of Eqs.~\eqref{eq:cuts1}--\eqref{eq:cuts2} and the additional CJV requirements of Eq.~\eqref{eq:cjv}--\eqref{eq:ptveto} at NLO accuracy and compare them to various \NLOPS{} setups.  Similar to the case without a CJV, we find that results do not change significantly when PS effects are taken into account, and that they are relatively stable with respect to PS settings. Only at very low and very high transverse momenta deviations between the global and the dipole recoil scheme become noticeable. 

\begin{figure}[t]%[htb]
	\centering
	\includegraphics[width=0.31\textwidth]{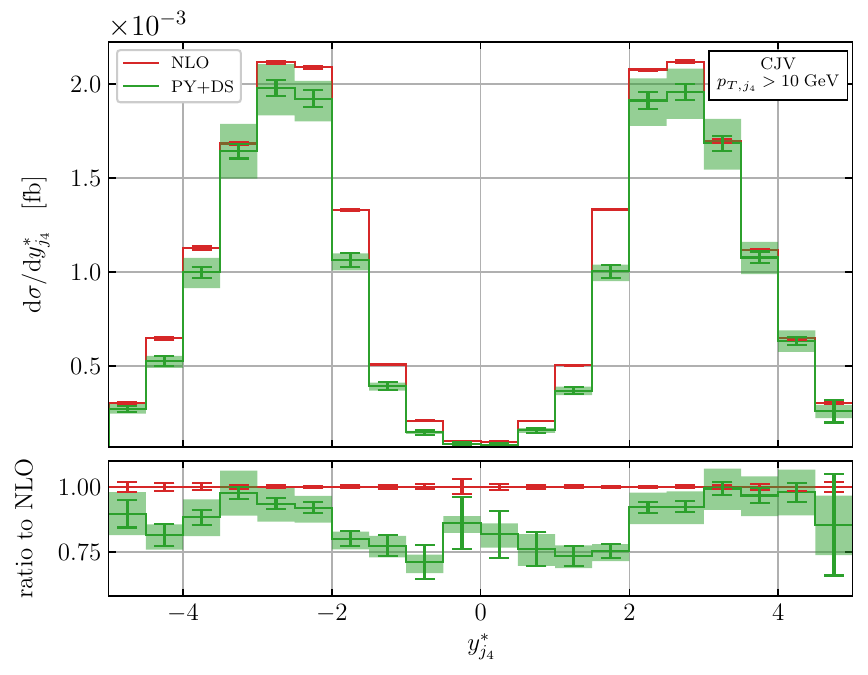}
	\includegraphics[width=0.31\textwidth]{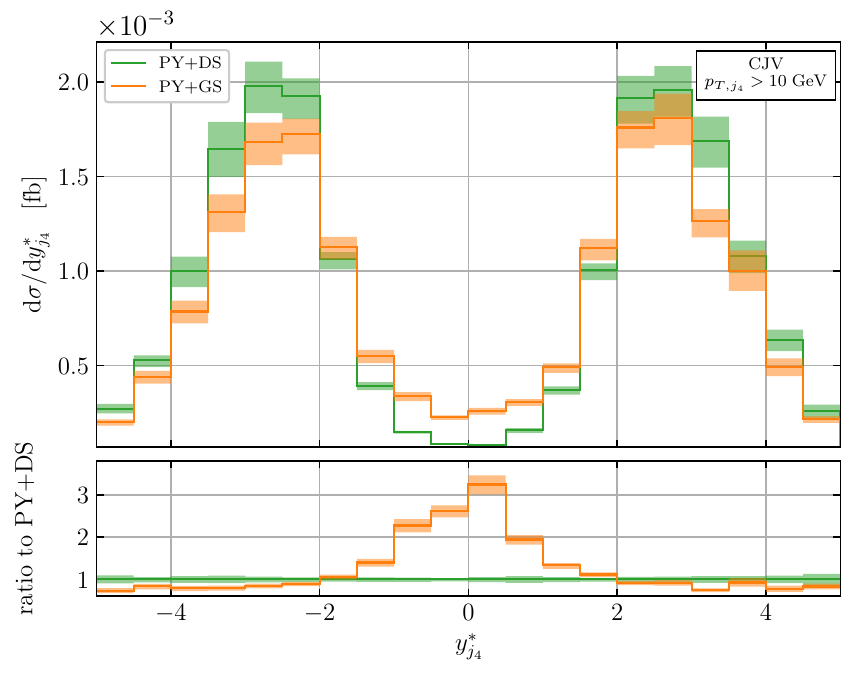}
	\includegraphics[width=0.31\textwidth]{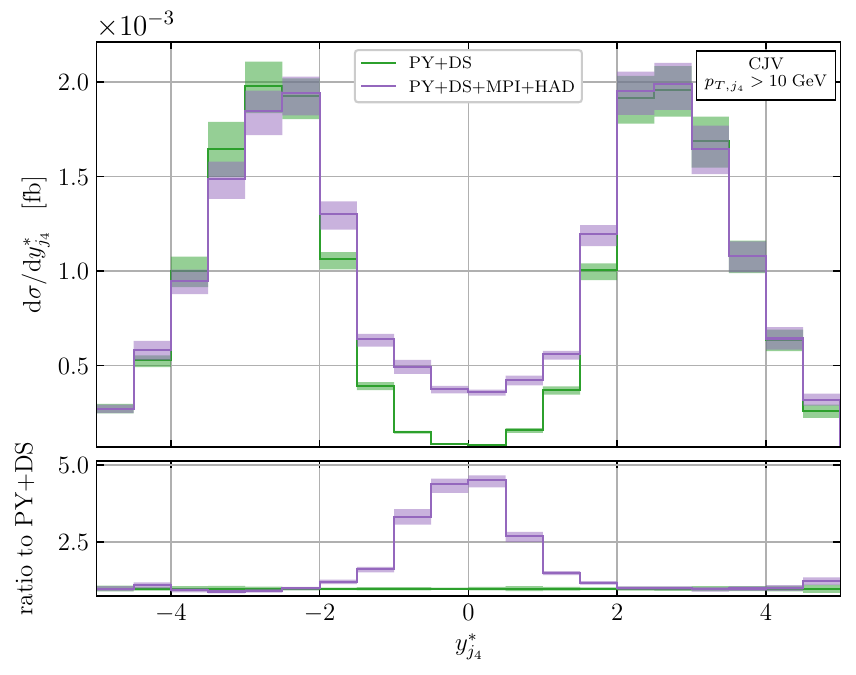}
	\caption{Similar to Fig.~\ref{fig:cjv-pt1}, but for the rapidity distributions of the fourth jet relative to the tagging jets and the additional cut of $p_{T,j_4}\geq 10~\gev$. }  
	\label{fig:cjv-ystar4}
\end{figure}

Let us now turn to distributions of the subleading jets. Figure~\ref{fig:cjv-ystar4} shows the rapidity distribution of the fourth jet relative to the tagging jet system that has been considered in Fig.~\ref{fig:NLO_vs_PY8-nontagjets} and Fig.~\ref{fig:PY8-comp} without a CJV. 
Note that for the investigation of CJV effects instead of \refeq{eq:cut-j4} we now impose the looser selection criterion of $p_{T,j_4}\geq 10~\gev$, since a stronger cut  would entirely remove radiation in the region of low $|y_{j_4}^*|$. Despite these slightly different settings the qualitative impact of PS effects on the NLO distribution is similar to the one observed without a CJV, see Fig.~\ref{fig:NLO_vs_PY8-nontagjets}. 
However, differences due to the PS setup are now significantly larger than without a CJV. Indeed, when the global recoil scheme is used the central region is filled out more noticeably than with the dipole recoil scheme, emphasizing the need to avoid the use of a global recoil scheme in the simulation of VBS processes. We also notice that MPI and hadronization effects tend to fill the region of low relative rapidity where the fourth jet is close to the center of the tagging-jet system. Radiation caused by such non-perturbative effects is typically soft and thus not affected by the CJV criterion removing events with jets of transverse momenta harder than 20~GeV.  

\begin{figure}[t]%[htb]
	\centering
	\includegraphics[width=0.55\textwidth]{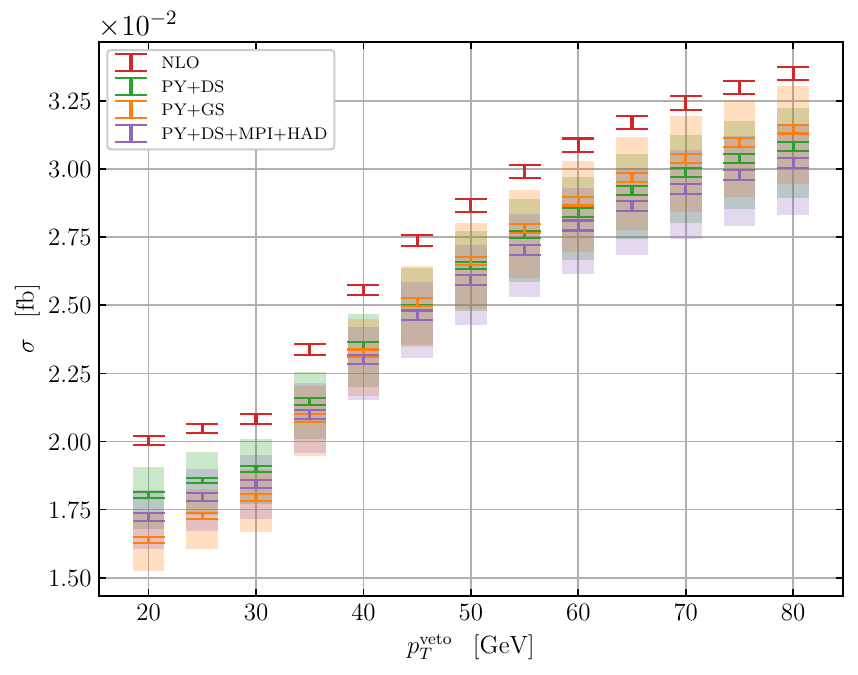}
	\caption{Cross section for $pp\to \wpwpjjj$ at the LHC as a function of $p_T^\mr{veto}$ within the cuts of Eqs.~\eqref{eq:cuts1}--\eqref{eq:cuts2} and Eq.~\eqref{eq:cjv} at NLO (red) and at \NLOPS{} accuracy using the \PYDS{} (green),  \PYGS{} (orange), and \PYMPI{} setups (purple).  
	The bands of the \NLOPS{} predictions indicate scale uncertainties obtained by a 7-point variation of $\muf$ and $\mur$.  
 }  
	\label{fig:cjv-xsec}
\end{figure}

Finally, let us consider the dependence of the $\wpwpjjj$ production cross section on the veto requirements in Fig.~\ref{fig:cjv-xsec}. 
While previously we have fixed the value of $\ptveto$ to 20~GeV, we now vary it up to values of 80~GeV where the CJV has a significantly smaller impact on the VBS cross section. 
We remind the reader that even before imposing a CJV, we observed a reduction of the NLO cross section by about 8.2\% because of PS effects. Similarly, after the application of a CJV the \NLOPS{} predictions obtained for various PS setups are slightly smaller than the corresponding NLO values while the different \NLOPS{} predictions overlap within their respective scale uncertainties.

% !TeX spellcheck = en_GB
%%%%%%%%%%%%%%%%%%%%%%%%%%%%%%%%%%%%%%%%%%%
\section{Summary and conclusions}\label{sec:conclusion}
%%%%%%%%%%%%%%%%%%%%%%%%%%%%%%%%%%%%%%%%%%%
%
In this article we presented a calculation for the EW on-shell production of two $W^+$ bosons and three jets in hadronic collisions making use of the VBS approximation. We computed the NLO-QCD corrections and 
matched them with PS programs making use of the \PBOXRES{}. 
Having full control on the kinematics of the tagging jets and the first subleading jet at NLO-QCD accuracy, and being able to describe the fourth jet with LO matrix elements, our implementation of the $\wpwpjjj$ production process provides unique means for the quantitative description of observables that are essential for the design of central-jet veto techniques. 

To illustrate the capabilities of our new Monte-Carlo program and  to explore the effect of QCD corrections and PS radiation on jet observables we performed a representative analysis of VBS-induced $\wpwpjjj$ production at the LHC with a centre-of-mass energy of 13.6~GeV. We found moderate NLO-QCD corrections for the cross section after selection cuts on the jets typical for VBS analyses, and showed that distributions of the tagging jets are only mildly modified by both NLO-QCD corrections and PS effects. Significant changes are observed for distributions of the sub-leading jets. In particular, we found that using a dipole recoil scheme is crucial when the NLO calculation is matched with \PYTHIA{}, in agreement with recent recommendations for the simulation of VBS processes in the literature.  

%%%%%%%%%%%%%%%%%%%%%%%%%%%%%%%%%%%%%%%%%%%
\section*{Acknowledgements}
%%%%%%%%%%%%%%%%%%%%%%%%%%%%%%%%%%%%%%%%%%%
We would like to thank A.~Denner and S.~Uccirati for their help with \recola{} and \recolatwo{} and valuable comments, and C.~Borschensky and C.~Schwan for helpful conversations.

Part of this work was performed on the high-performance computing resource 
bwForCluster NEMO with support by the state of Baden-W\"urttemberg through bwHPC and the German Research Foundation (DFG) through grant no INST 39/963-1 FUGG.

\bibliography{VBS3j}%{}

\end{document}